%% file: Galton_model_FINAL2.tex
\documentclass[aps,amssymb,amsmath,pra,twocolumn,showpacs,nosuperscriptaddress,longbibliography]{revtex4-2}

\usepackage{graphicx}
\usepackage{dcolumn}
\usepackage{bm}
\usepackage{subfigure}
\usepackage{color}
\usepackage{blkarray}

\usepackage{amssymb,amsmath,amsfonts,latexsym,graphicx,verbatim}

\usepackage[margin=0.55in]{geometry}

\usepackage[english]{babel}
\usepackage{times}
\usepackage{latexsym}
\usepackage{fancyhdr}
\usepackage{float}
\usepackage{dsfont}
\usepackage{afterpage}
\usepackage{enumitem}
\setlist[itemize]{noitemsep, topsep=0pt}
\usepackage{graphicx,subfigure}

\usepackage{listings}
\usepackage{multirow}
\usepackage[table]{xcolor}
\usepackage{enumitem}

\usepackage{bbm}
\usepackage{upgreek}

 \usepackage{nicematrix}
\NiceMatrixOptions{
code-for-first-row = \color{blue} ,
code-for-last-row = \color{blue} ,
code-for-first-col = \color{blue} ,
code-for-last-col = \color{blue},
}

\definecolor{Ablue}{rgb}{0.96,0.24,0.00}

\definecolor{Abluetitle}{rgb}{0.,0.24,0.51}

\definecolor{orange}{rgb}{0.96,0.24,0.00}

\definecolor{darkred}{rgb}{0.55, 0.0, 0.0}

\definecolor{Gray}{gray}{0.85}
\definecolor{LightCyan}{rgb}{0.88,1,1}
\definecolor{darksalmon}{rgb}{0.91, 0.59, 0.48}
\definecolor{maroon}{cmyk}{0,0.87,0.68,0.32}

\definecolor{mustard}{rgb}{1.0, 0.86, 0.35}
\newcolumntype{a}{>{\columncolor{Gray}}c}
\newcolumntype{b}{>{\columncolor{white}}c}

\usepackage{tikz}
\newcommand*\circled[1]{\tikz[baseline=(char.base)]{
            \node[shape=circle,draw,inner sep=2pt] (char) {#1};}}

\usepackage{array}
\newcolumntype{L}[1]{>{\raggedright\let\newline\\\arraybackslash\hspace{0pt}}m{#1}}
\newcolumntype{C}[1]{>{\centering\let\newline\\\arraybackslash\hspace{0pt}}m{#1}}
\newcolumntype{R}[1]{>{\raggedleft\let\newline\\\arraybackslash\hspace{0pt}}m{#1}}

\usepackage[colorlinks=true , citecolor=blue,urlcolor=blue]{hyperref}

\addto\captionsenglish{\renewcommand{\figurename}{Fig.}}

\input{Commands3.tex}

\newcommand{\affA}{Department of Chemistry, University of California Berkeley,  California 94720, USA.}

\begin{document}
\title{\T{\I{``Galton board''} nuclear hyperpolarization}}
\author{Moniish Elanchezhian}\affiliation{\affA}
\author{Arjun Pillai}\affiliation{\affA}
\author{Teemu Virtanen}\affiliation{\affA}
\author{Ashok Ajoy}\email{ashokaj@berkeley.edu}\affiliation{\affA}

\begin{abstract}
We consider the problem of determining the spectrum of an electronic spin via polarization transfer to coupled nuclear spins and their subsequent readout. This suggests applications for employing dynamic nuclear polarization (DNP) for \I{“ESR-via-NMR”}. In this paper, we describe the theoretical basis for this process by developing a model for the evolution dynamics of the coupled electron-nuclear system through a cascade of Landau-Zener anti-crossings (LZ-LACs). We develop a method to map these traversals to the operation of an equivalent \I{``Galton board’’}. Here, LZ-LAC points serve as analogues to Galton board “pegs”, upon interacting with which the nuclear populations redistribute. The developed hyperpolarization then tracks the local electronic density of states. We show that this approach yields an intuitive and analytically tractable solution of the polarization transfer dynamics, including when DNP is carried out at the \I{wing} of a homogeneously broadened electronic spectral line. We apply this approach to a model system comprised of a Nitrogen Vacancy (NV) center electron in diamond, hyperfine coupled to $N$ neighboring $\Cs$ nuclear spins, and discuss applications for nuclear-spin interrogated NV center magnetometry. More broadly, the methodology of \I{``one-to-many’’} electron-to-nuclear spectral mapping developed here suggests interesting applications in quantum memories and sensing, as well as wider applications in modeling DNP processes in the multiple nuclear spin limit.
\end{abstract}

\maketitle

 \vspace{-2mm}
\section{Introduction}
 \vspace{-1mm}
The central spin model is the physical description underlying a wide class of quantum mechanical systems relevant in quantum information processing~\cite{Arenz14,Witzel10}, sensing~\cite{Degen17,Taminiau12} and DNP~\cite{Hovav10,Hovav12}. This includes solid-state qubits or sensors constructed out of electronic defects and donors in materials such as silicon~\cite{Morello10,Pla12} and diamond~\cite{Jelezko06,Taylor08}, and in the context of DNP, radical species embedded within nuclear spins of analytes of interest~\cite{Maly08,Hu07}. These systems are typically comprised of \I{dilute} electronic spin centers interacting with a several-fold larger nuclear spin bath (schematically shown in \zfr{fig1}A). Assuming weak inter-electron couplings, the electronic spectrum is dominated by hyperfine couplings to the nuclei (\zfr{fig1}B). In this paper, we examine whether this homogeneously broadened spectrum can be extracted via interrogation of the nuclear spin bath alone.  We discuss how this might allow the possibility for \I{“ESR-via-NMR”} in some restricted settings. Such a method would form a complement to other techniques such as dynamical decoupling spectroscopy~\cite{Alvarez11,Bar-Gill12}. 

Consider the scenario in \zfr{fig1}A describing $N$ nuclear spins ($n$) coupled to an electronic spin ($e$) in a magnetic field $B_0$, while ignoring internuclear interactions. \zfr{fig1}B depicts the resulting energy levels in two electronic manifolds ($m_s{=}\{0,\R{+}1\}$), assuming a spin-1 electron and $N{=}3$ spin-1/2 nuclei. We refer to the electronic spectral density of states (DOS) as the number of electronic levels per unit frequency in a particular manifold (see \zfr{fig2}B).  Our approach utilizes a DNP protocol~\cite{Ajoy17,Ajoy18,Ajoy20} (\zfr{fig2}A) for selective hyperpolarization injection from the central spin to the nuclear bath. We show that the extent of polarization transferred reflects the underlying electronic DOS (see \zfr{fig2}B-C), and therefore reports on the electronic spectrum (see accompanying paper~\cite{Pillai21}). This exploits special features of polarization transfer when employed at the wing of the electronic spectral line.  Assume that the electron is polarized in one of the manifolds, here $m_s{=}0$, and the nuclear spins start in a completely mixed state $\T{1}_N$. DNP is excited by the application of frequency-swept irradiation (\zfr{fig2}A) across a narrow window $\mB$ whose size is much narrower than the spectral linewidth (\zfr{fig2}B).  The magnitude of hyperpolarization is then found to \I{track} the DOS being swept over, as demonstrated by the companion paper to this article~\cite{Pillai21} (see \zfr{fig2}C). 

 \begin{figure}[H]
  \centering
  \includegraphics[width=0.49\textwidth]{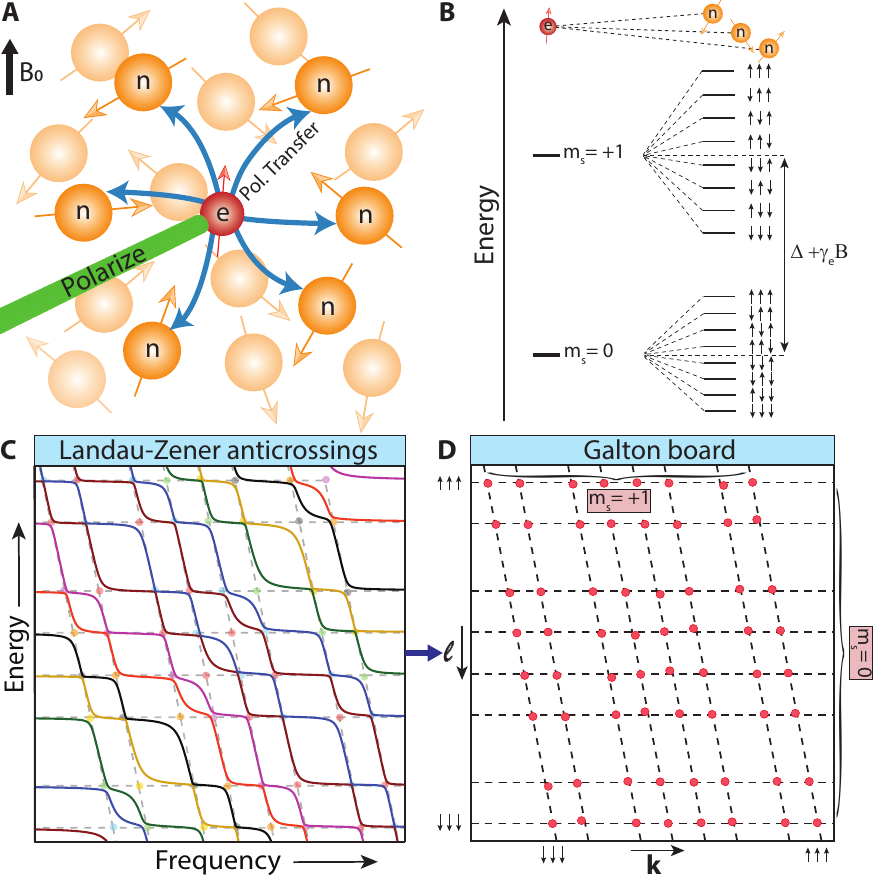}
  \caption{\T{Principle. } (A) \I{Central spin model} consisting of an electron ($e$) coupled to $N$ nuclear spins ($n$) in a magnetic field $B_0$, ignoring internuclear couplings. We focus here on an NV center electron coupled to $\Cs$ nuclei. The NV electron  is polarized by optical pumping (green), and polarization is transferred (blue arrows) to $\Cs$ nuclei by means of microwave (MW) irradiation (see \zfr{fig2}). (B) \I{Lab frame} energy levels of the $e$-$n$ system for $N{=}3$, showing a wide separation of nuclear levels arranged by electronic manifold.  (C) \I{LZ anti-crossings}. In the rotating frame of the applied MWs, the levels in (B) transform to a cascade of LZ anti-crossings (\I{solid lines}) between nuclear states in alternate electronic manifolds. Dashed lines show crossings corresponding to the diagonal Hamiltonian (see \zfr{fig5}). Dots refer to the crossing points. (D) \I{Equivalent Galton Board}. Isolating the crossing points allows us to abstract the system in terms of a Galton board (see \zfr{fig3}) where the crossing points (red) arrange in the checkerboard $\mI_{k,\ell}$ in a 2D plot of energy and frequency (labeled $k$ and $\ell$ respectively). The two electronic manifolds are highlighted for clarity.}
\zfl{fig1}
\end{figure}

A primary focus of this manuscript is to model this surprising result (detailed in Ref. \cite{Pillai21}) via the corresponding polarization transfer dynamics to $N$ nuclei. \zfr{fig1}C-D intuitively capture the physics of the DNP process; lab frame energy levels in \zfr{fig1}B are transformed to a series of $2^{2N}$ cascaded LZ  level anti-crossings (\I{``LZ-LACs"}) in the rotating frame (see \zfr{fig1}C), wherein ground state electronic levels successively anti-cross with their excited state counterparts. Here, we describe a mechanism to model the dynamics of traversals through such an \I{“LZ-cascade”}. Building an analogy, we show that the process can be mapped onto the operation of a \I{“Galton board”}~\cite{Lue93,Bouwmeester99,Chernov07} (\zfr{fig3}). Here, we isolate the position of the LZ-LACs (points in \zfr{fig1}C) and lay them out in a 2D space of energy and frequency (axes labeled $k$ and $\ell$) as in \zfr{fig1}D, where the horizontal and vertical lines refer to the two electronic manifolds (highlighted). The resulting \I{checkerboard} pattern in \zfr{fig1}D, which we refer to as $\mI_{k\ell}$, makes concrete the analogy to the Galton board (\zfr{fig3}) — the LZ points form \I{``pegs''}, upon interacting with which, under frequency swept drive, the nuclear populations redistribute amongst the various levels within an electronic manifold. We show that this provides a simple way to track the population changes at every time instant, ultimately capturing the essential physics of the experiments. 

\begin{figure}[t]
  \centering
 \includegraphics[width=0.49\textwidth]{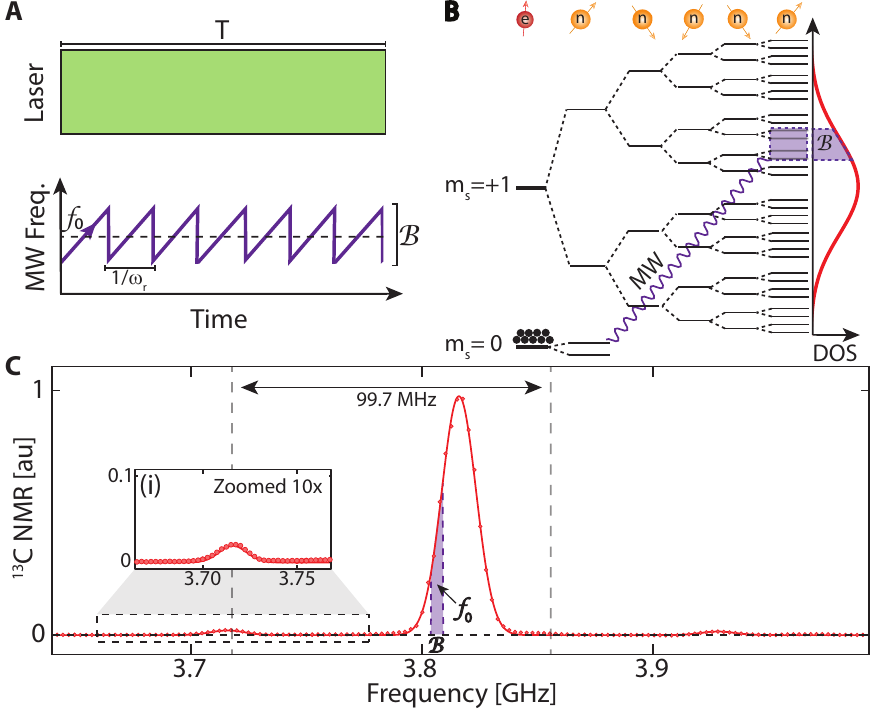}
  \caption{\T{Experimental motivation.} (A) \I{DNP protocol}. NV${\rt}\Cs$ polarization transfer is driven by chirped MWs performed in narrow frequency windows (centered at $f_0$ with bandwidth $\mB$) under continuous laser illumination~\cite{Ajoy17, Ajoy18} for period $T$. (B) \I{Schematic of hyperpolarization} as it occurs at the wing of the electronic spectral line. Representative levels showing eigenstates of the diagonal Hamiltonian in the upper electronic manifold with increasing numbers of nuclear spins. Starting with polarization at $m_s{=}0$ (black balls), chirped MWs (curly purple line) drive transitions to the $m_s{=}$+1 manifold in the small $\mB$ window (purple box) as shown. Solid red line denotes the electronic DOS in the upper manifold. (C) \I{Experimental data} on a single-crystal sample with natural abundance $\Cs$ and $\app$1ppm NV concentration. DNP here is performed at $B_0{\app}$33.6mT in $\mB$=10MHz windows (purple box) whose center is scanned in frequency. $\Cs$ NMR signal here is measured at 7T. $\Cs$ hyperpolarization signal \I{tracks} the local electronic DOS (see also companion paper~\cite{Pillai21}). Solid line is a Gaussian fit.  \I{Inset (i)}: Zoom in to satellites, hyperfine shifted by $\app$99.7MHz corresponding to NV centers with first-shell $\Cs$ nuclei. }
\zfl{fig2}
\end{figure}

\begin{figure}[t]
  \centering
 \includegraphics[width=0.49\textwidth]{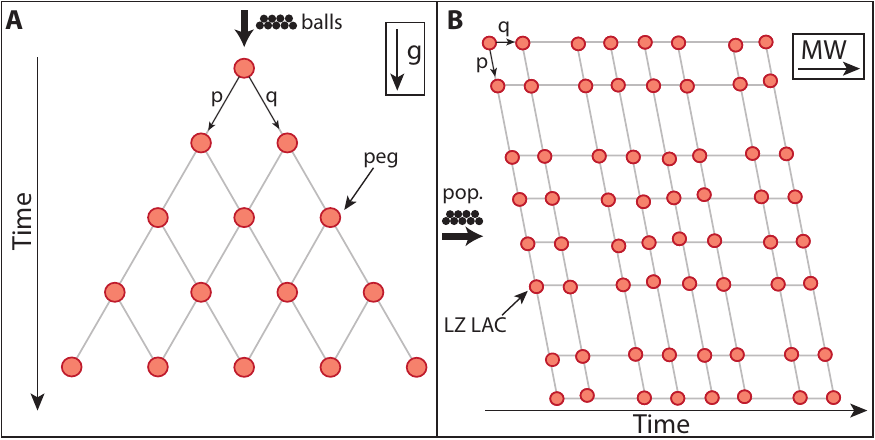}
  \caption{\T{Galton Board Schematic.} (A) \I{Classical ``Galton Board"} consisting of pegs placed in a pyramidal fashion. Balls striking pegs bifurcate left or right with probabilities $p$ or $q$ respectively. Driving force is provided by gravity (marked in box). (B) \I{LZ Galton Board}. Analogous Galton board (see \zfr{fig1}D) formed in our system where pegs are the LZ-LAC points. Shown is the case when $N{=}3$ (\zfr{fig1}C). Balls are nuclear populations, which bifurcate down or right. Driving force is provided by the applied MW sweep (see \zfr{fig2}). }
\zfl{fig3}
\end{figure}

 \vspace{-2mm}
\subsection{Experimental motivation for this work}
 \vspace{-1mm}
The motivation for this work is the experiment shown in \zfr{fig2}C and in the companion paper~\cite{Pillai21}, performed in a model system in diamond.  Herein, the electron in \zfr{fig1}A is a spin-1 NV defect center, surrounded by multiple $\Cs$ nuclear spins. Attractive spin-optical properties allow NV electronic polarization to the $m_s{=}0$ state via optical pumping at room temperature. Simultaneously, long $\Cs$ lifetimes permit the ability to interrogate them for minute long periods~\cite{Beatrez21}, providing high signal-to-noise means to read them out. 

DNP is applied at low field ($B_0{\app}$1-70mT) using a MW frequency chirp over a narrow window $\mB$ centered at frequency $f_0$ and at a repetition rate $\xo_r$ (\zfr{fig2}A)~\cite{Ajoy17, Ajoy20}. \zfr{fig2}B schematically represents this (purple window) on the NV spectrum, hyperfine-broadened by coupling to a few $\Cs$ nuclei. The spectral DOS is also schematically drawn in on the right side of \zfr{fig2}B (red line). The resulting $\Cs$ polarization is measured at high field (7T). Interestingly, we observe (see \zfr{fig2}C) that as $f_0$ is scanned across in frequency, the $\Cs$ NMR signals (points) closely track the underlying NV ESR spectral DOS. The satellites in \zfr{fig2}C, shifted ${\approx}$99.7MHz with respect to the spectral center, correspond to NV centers with a $\Cs$ nucleus in their first shell~\cite{Rao16}. A more detailed exposition of these experiments is presented in the accompanying paper~\cite{Pillai21}. Here we also study the scaling of the obtained linewidth with the size of the sweep window $\mB$ and the influence of the MW sweep direction. We find that the width of the DNP derived spectrum (as in \zfr{fig2}C) saturates at ${\app}$2MHz on the lower end, closely resembling the intrinsic ESR linewidth. These experiments suggest that \I{“ESR-via-NMR”} may be feasible in such systems~\cite{Pillai21}.

While some aspects of the physics here are easy to intuit — sweeping over a part of the spectrum in \zfr{fig2}C where there is negligible electronic DOS intensity leads to no hyperpolarization — it is still mysterious that narrow sweeps applied to the wing of a spectral line should produce a $\Cs$ polarization that seemingly tracks the local spectral DOS. We note, however, that the fact of DNP enhancement amplitudes reflecting the underlying electronic spectrum, at least \I{qualitatively}, is common knowledge~\cite{Can15,Becerra93,Mentink17,Michaelis13}. It is widely employed in experiments optimizing DNP levels and to characterize the effectiveness of a radical species in yielding hyperpolarization~\cite{Leavesley18}. However, the difference in the present work is that we seek to make this connection more \I{quantitative}, exploiting the DNP levels here to \I{“sense”} the ESR spectrum. Ultimately, this also portends applications in magnetometry, as explored in Ref. ~\cite{Pillai21} exploiting the ability to readout the nuclear spins with high signal-to-noise employing spin-locking techniques~\cite{Beatrez21}.

We benefit in this endeavor from special features of the employed DNP mechanism~\cite{Ajoy17, Ajoy20} (\zfr{fig2}A). Every part of the spectrum produces hyperpolarization, but of an identical sign, as evidenced by \zfr{fig2}C. This is in contrast to common DNP techniques (e.g. solid effect~\cite{Hovav10}) where the DNP sign is inverted at the red or blue shifted side of the spectral center. The latter can cause cancellation in the DNP enhancements between different parts of the spectrum, especially at low fields or for inhomogeneously broadened spectra, making determining the underlying spectrum non-trivial~\cite{Ramanathan08}.

It is evident that a full description of the DNP mechanism yielding the results at the \I{wing} of the spectrum in \zfr{fig2}C should include the physics of polarization dynamics when the electron is coupled to $N$ nuclear spins.  In that respect, our work contributes to strategies for the analytic descriptions of DNP in the large $N$ limit,  extending common approaches that focus primarily on $e$-$n$ or $e$-$e$-$n$ systems (e.g. solid and cross effects~\cite{Hovav10,Thurber12,Hovav12}), and where larger nuclear networks are evaluated fully numerically~\cite{Hovav10,Butler09,Mentink15}. Indeed, previous calculations of the DNP technique~\cite{Ajoy17} employed here were also carried out only in the single $e$-$n$ limit~\cite{Zangara18}. Instead in this paper, relying on a set of simplifying assumptions and a mapping to a Galton board (\zfr{fig3}), we show that a mechanistic description with $N$ coupled nuclei as in \zfr{fig1}A is tractable, and can lead to experimentally verifiable predictions.

 \vspace{-2mm}
\subsection{Summary of approach}
 \vspace{-1mm}
\zfr{fig1}C-D and \zfr{fig4} summarize the basic theoretical approach. The problem of solving the hyperpolarization dynamics becomes that of tracking the populations in a system as they encounter a cascade of LZ-LACs under a MW sweep. The exponentially scaling number of anti-crossings (${\propto}2^{2N}$) makes this problem challenging; however with some approximations (detailed below), the dynamics can be restricted to that of populations at the LZ points alone (\zfr{fig1}D) .

Our solution entails building an analogy to the operation of a Galton board. Consider the LZ checkerboard in \zfr{fig4}. At $t{=}0$, the nuclear populations are equally distributed along the levels of the $m_s{=}0$ manifold. Upon encountering an LZ point, however, they are redistributed with a finite probability either \I{“right”} or \I{“down”} on the checkerboard $\mI_{k,\ell}$ with the probability defined by the size of the local LZ energy gap. Tracking the population dynamics can now be handled with a transfer matrix formalism for any specific trajectory traversed through the LZ checkerboard.  If $\T{\I{v}}$ defines a path through ${n}$ points in the 2D checkerboard (black arrow in \zfr{fig4}) with coordinates $v{=}\{v_j\} {=} \{(k_1,\ell_1), (k_2,\ell_2),\cdots,(k_j,\ell_j)\cdots (k_{n},\ell_{n})\}$ (see \zfr{fig4}B), and $\T{p}_i$ ($\T{p}_f$) are column vectors denoting nuclear populations at the initial (final) LZ point, traversals can be evaluated analogous to the operation of a Galton machine,
\beq
\T{p}_f {=} \T{T}_{v_n}\T{M}_{(v_{n},v_{n-1})}\cdots\T{T}_{v_j}\T{M}_{(v_{j},v_{j-1})}\T{T}_{v_2}\T{M}_{(v_2,v_{1})}\T{T}_{v_1}\T{M}_{r}\cdot \T{p}_i\non
\eeq
where $\T{T}_{v_j}$ is a \I{“redistribution”} operator at the LZ point $v_j$, and $\T{M}_{(v_{n},v_{n-1})}$ and $\T{M}_{r}$ are \I{“walk”} operators connecting the nearest neighbor points that define the path $\T{\I{v}}$. Through this simple map, we demonstrate the origin of \I{bias} in the traversals which is responsible for nuclear hyperpolarization. We then show that DNP carried out in small windows $\mB$ of the electronic spectrum occurs such that the hyperpolarization follows the electronic DOS, a reflection of the experimental data in \zfr{fig2}C.

\begin{figure}[t]
  \centering
 \includegraphics[width=0.49\textwidth]{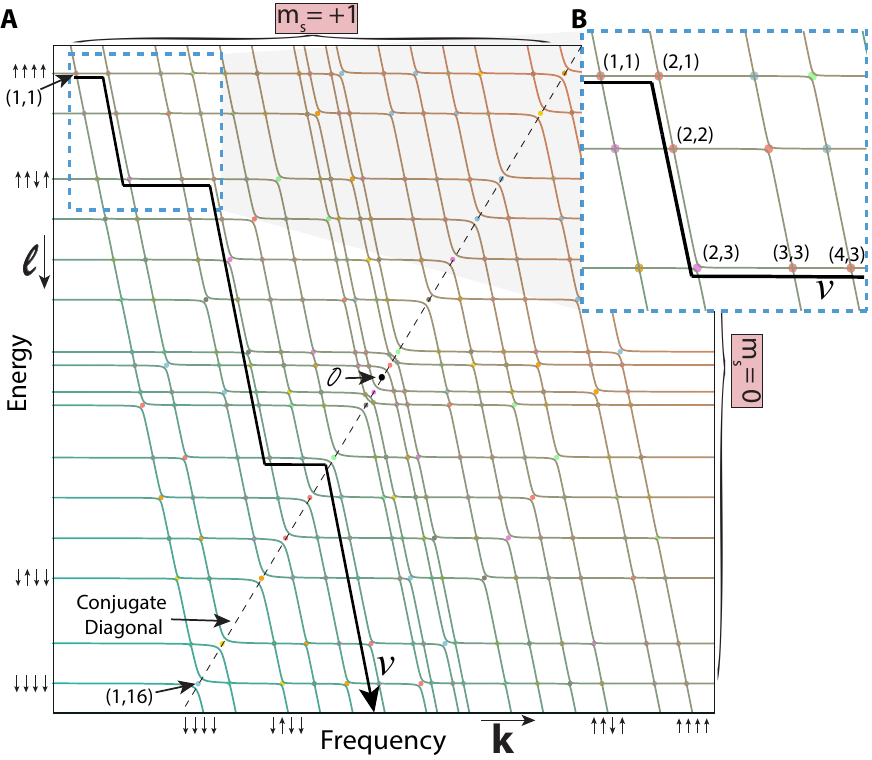}
  \caption{\T{Galton board of Landau-Zener LACs}. (A) Panel shows the Landau-Zener LACs for a representative case of an NV center coupled to $N{=}4$ $\Cs$ nuclei and positive $A_j^{\pll}$. Representative nuclear states are marked. The LZ-LACs (dots) form the $\mI_{k,\ell}$ checkerboard as in \zfr{fig1}D, where coordinates mark their individual positions. Axes $\{k,\ell\}$ refer to energy and frequency respectively. Center of the checkerboard is marked $\mO$ and the conjugate diagonal (\I{dashed line}) refers to positions $\mI_{k,2^N-k+1}$. Eigenstates corresponding to the diagonal Hamiltonian $\mH_{\R{diag}}$ are labeled (see \zfr{fig5}). At $t{=}0$, populations start equally distributed amongst the nuclear states restricted to the $m_s{=}0$ manifold. A MW sweep causes the system to encounter the LZ-LACs in a sequential manner. The black arrow shows an exemplary  trajectory $\T{\I{v}}$ through the checkerboard. (B) \I{Zoom-in} to a section (blue dashed box) of the checkerboard showing specific vertices $v_j$ forming the trajectory $\T{\I{v}}$. Nuclear populations are redistributed at every LZ-LAC similar to the operation of a Galton board (\zfr{fig3}). }
\zfl{fig4}
\end{figure}

\begin{figure*}[t]
  \centering
  {\includegraphics[width=0.95\textwidth]{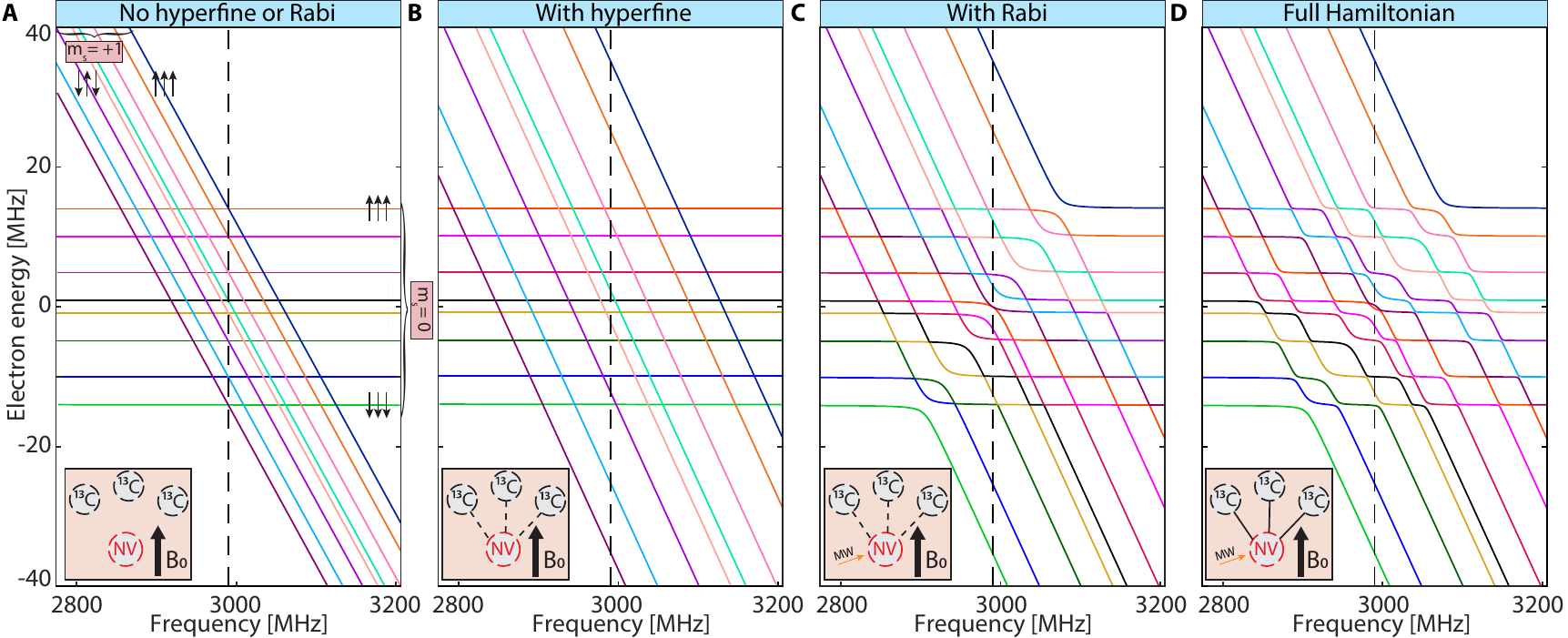}}
  \caption{\T{Emergence of cascaded Landau-Zener LACs} described in a systematic manner for an NV center coupled to $N{=}3$ $\Cs$ nuclei in a magnetic field $B_{0}$ (see \zfr{fig1}A) using the model in \zr{eq9}. We focus on the $m_s{=}\{0,\R{+}1\}$ electronic manifolds, and introduce Hamiltonian terms hierarchically to illustrate the physical origin of the LZ-LACs (\I{shaded insets}). (A) Considering only the diagonal Hamiltonian $\mH_{\R{diag}}$ (\zr{eq5}) with $A_j^{\pll}{=}0$, and (B) with finite $A_j^{\pll}$. There are $2^{2N}$ crossing points between nuclear states in the two electronic manifolds, labeled here by the corresponding eigenstates (shown are representative states). (C) Including now the Rabi frequency term $\xO_e$ from $\mH_{\R{non-diag}}$ (\zr{eq6}) leads to LZ-LACs between states on the conjugate diagonal (see \zfr{fig4}) involving a direct electronic spin flip, $\ket{0,\mS}{\ltrt}\ket{\R{+}1,\mS}$ for any nuclear state $\mS$. (D) Finally, considering the full Hamiltonian $\mH_{\R{tot}}$ (\zr{eq4}), LZ-LACs appear at each crossing point. The corresponding energy gaps display the native hierarchy based on the total number of spin flips they involve (see App. A). }
\zfl{fig5}
\end{figure*}

 \vspace{-2mm}
\section{System of Cascaded Landau-Zener LACs}
\zsl{cascade}
 \vspace{-2mm}

\subsection{Central spin system and hyperpolarization protocol}
 \vspace{-1mm}
The NV-$\Cs$ system is considered here as a model of a hybrid $e$-$n$ system that captures two commonly encountered features reflected in \zfr{fig1}A:
\benum[noitemsep,topsep=0pt,label=(\roman*)]
\item	 the electron concentration is orders of magnitude more dilute than that of the nuclear spins, and thus inter-electron couplings can be neglected, and,
\item	nuclear $T_{1n}$ far exceeds that of the electron, and polarization can be made to accumulate in the nuclei through repeated polarization transfer from the electron.
\eenum

In our samples (e.g as employed in \zfr{fig2}C), the NV centers are surrounded by an average of ${\sim}10^4$ $\Cs$ nuclei, and inter-electron spacings are $\expec{r_e}{\app}$12nm~\cite{Ajoy19relax}. If $w_L$ defines the rate of NV electronic polarization buildup, the electron polarization follows as $P_e(\qt) \app 1-\exp(-\qt w_L)$. At sufficiently high optical powers, $w_L$ is set by the intensity of the laser illumination applied. At low powers however, $w_L{\app} T_{1e}^{-1}$~\cite{Ajoy20}; the electron polarization rate is then set by just the thermal regeneration rate.

 \vspace{-2mm}
\subsection{NV-$\Cs$ System Hamiltonian}
 \vspace{-1mm}
The cascade of LZ-LACs as in \zfr{fig1}C and \zfr{fig4} originates from diagonalizing the \I{instantaneous} system Hamiltonian, 
\beq
\mH_{\R{tot}}(t) = \mH + \mH_{\R{MW}}(t) ,
\zl{eq1}
\eeq
at every step of the applied frequency sweep. The two terms here refer to the spin Hamiltonian and applied chirped MW control respectively. The latter has the form, 
\beq 
\mH_{\R{MW}}(t) = \xO_e S_x\cos(\xo_{\R{MW}} t) 
\zl{eq2} 
\eeq 
where $S_x$ is a spin-1 Pauli operator and $\xO_e$ is the electronic Rabi frequency. The instantaneous MW frequency, $\xo_{\R{MW}}(t)$, is of the form, 
\beq
\xo_{\R{MW}}(t) = \mB\xo_r t+{f_0}-\mB/2
\zl{eq3}
\eeq 	
where $\xo_r$ is the sweep repetition rate.  We will focus here on a scenario as in \zfr{fig1}A, considering a system of $N$ $\Cs$ nuclear spins directly hyperfine-coupled to an NV center with hyperfine strength $\|A_j\|$, where $j=1, {\cdots}, N$. We ignore dipolar couplings between the $\Cs$ nuclei, focusing instead on a direct NV${\rt}\Cs$ polarization transfer process.  In the low-field DNP regime,  the bare $\Cs$ Larmor frequency is smaller than or comparable to the strength of the hyperfine couplings, $\xo_L{=}\xg_nB_0{\lesssim} \|A_j\|$~\cite{Ajoy17}, where $\xg_n{=}1.07$kHz/G is the nuclear magnetogyric ratio.

Now considering the spin Hamiltonian $\mH$ for a system of $N$ $\Cs$ nuclear spins hyperfine-coupled to the NV center as in \zfr{fig1}A, it is convenient to consider the \I{nuclear} Hamiltonian selectively in each NV manifold. We focus here just on the two-level system formed by the $m_s{=}\{0,\R{+}1\}$ manifolds (see \zfr{fig1}B). We assume for simplicity that $\T{B}_0$ is aligned with the NV axis, a direction we label $\hat{z}$. We separate the hyperfine field into its longitudinal and transverse components, $\T{A}_j{=} A_{j}^{\pll}\hat{z} + A_{j}^{\pp}\hat{\pp}$, where $\hat{\pp}$ is a unit-vector in the $xy$ plane.   Some comments are worth noting:
\benum[label=(\roman*)]
\item Experiments (e.g. \zfr{fig2}C) are performed in the ensemble limit and thus, to a good approximation, there is an equal probability of spins with positive or negative $A^{\pll}$ couplings.
\item	The $m_s{=}0$ NV state is ``non-magnetic'', and the nuclear spins are predominantly quantized here in the Zeeman field $\T{B}_0$ with frequency $\xo_L{=}\xg_nB_0$. In reality, there is a weak additional second-order coupling mediated by the hyperfine term, leading to an effective nuclear frequency in the $m_s{=}0$ manifold, $\xo^{(0)}_j {\app} \xo_L {+} \fr{\xg_e B_0 A_{j}^{\pp}}{\xD}$, where $\xD$=2.87GHz is the NV center zero field splitting. The nuclear Hamiltonian in the $m_s{=}0$ manifold is then of the form, $\mH_{n}^{(0)}{=}\sum_{j=1}^{N}\xo_j^{(0)}I_{zj}$,where $I$ are spin-1/2 Pauli operators. 
\item In the $m_s{=}$+1 manifold, on the other hand, there is a combined action of the Zeeman and hyperfine fields, $\xo^{(1)}_j{=} \sqrt{(\xo_L + A_{j}^{\pll})^2 + (A_{j}^{\pp})^2}$. Indeed, the quantization axis $\hat{z'_j}$ here need not be collinear with $\T{B}_0$; in general, $\hat{z’_j} {=} \hat{z}\cos\xph_j {+} \hat{x}\sin\xph_j$, where the angle $\xph_j{=}\tan^{-1}\lsb A_{j}^{\pp}/(\xo_L + A_{j}^{\pll})\rsb$. The $m_s{=}$+1 nuclear Hamiltonian is then of the form, $\mH_{n}^{(1)}{=}\sum_{j=1}^{N}\xo_j^{(1)}I_{z'j}$, where $I_{z'j} {=} I_{zj}\cos\xph_j + I_{xj}\sin\xph_j$.   
\eenum

Going into a rotating frame with respect to $S_z^2$ at frequency $\xo_{\R{MW}}$, gives the net system Hamiltonian in \zr{eq1},
\begin{widetext}
\beq
\mH_{\R{tot}}(\xo_{\R{MW}}) = (\xD - \xo_{\R{MW}})S_z^2 + \xg_e B_0 S_z + \xO_e S_x + \sum_{j=1}^{N}\lsb \xo_j^{(0)}\mP_0I_{zj} + \xo_j^{(1)}\mP_1I_{z'j}\rsb\:,
\zl{eq4}
\eeq
\end{widetext}
where $\mP_0=\mat{1}{0}{0}{0}\otimes\T{1}_N\:;\: \mP_1=\mat{0}{0}{0}{1}\otimes\T{1}_N$, are the respective projection operators to the NV manifolds, and $\T{1}_N$ is a unit operator on the nuclear spins.

 \vspace{-2mm}
\subsection{Eigenstates and eigenenergies}
 \vspace{-1mm}
Plotted with respect to the instantaneous frequency $\xo_{\R{MW}}$, the eigenvalues of $\mH_{\R{tot}}$ in \zr{eq4} undergo a sequence of cascaded LZ-LACs (as in \zfr{fig1}C and \zfr{fig4}). To determine their positions, our strategy will be to first calculate the crossing points, focusing only on the diagonal Hamiltonian, and subsequently promoting them to anti-crossings by including off-diagonal terms (see \zfr{fig5}).  To this end, consider first separating $\mH_{\R{tot}}$ into its diagonal and off-diagonal parts, $\mH_{\R{tot}}(\xo_{\R{MW}}) = \mH_{\R{diag}}(\xo_{\R{MW}}) + \mH_{\R{non-diag}}$, where,
\begin{widetext}
\beq
\mH_{\R{diag}}(\xo_{\R{MW}}) = (\xD - \xo_{\R{MW}})S_z^2 + \xg_e B_0 S_z + \sum_{j=1}^{N}\lsb \xo_j^{(0)}\mP_0I_{zj} + \xo_j^{(1)}\mP_1I_{zj}\cos\xph_j\rsb\:\\
\zl{eq5}
\eeq
\vspace{-6mm}
\beq
\mH_{\R{non-diag}} = \xO_e S_x + \sum_{j=1}^{N} \xo_j^{(1)}\mP_1I_{xj}\sin\xph_j\:
\zl{eq6}
\eeq
\end{widetext}
The latter encapsulates contributions from $\{\xO_e,A_j^{\pp}\}$ terms.  Considering only $\mH_{\R{diag}}$ allows us to label the eigenstates (\zfr{fig1}D), and identify points at which LZ-LACs will ultimately arise (see also \zfr{fig4} and \zfr{fig5}).  Specifically, in \zfr{fig5}A-D, we show the systematic emergence of the LZ-LACs in a series of steps highlighted in the insets —
\benum[label=(\roman*)]
\item	considering $\mH_{\R{diag}}$, but with only the Zeeman terms of the electron and nuclear spins (\zfr{fig5}A);
\item	considering $\mH_{\R{diag}}$, but including the hyperfine terms (\zfr{fig5}B) results in the crossing points that closely define the checkerboard $\mI_{k,\ell}$ as in \zfr{fig1}D;
\item	including the Rabi term from $\mH_{\R{non-diag}}$ (see \zfr{fig5}C) results in a set of anti-crossings at many former crossing points, here corresponding to levels where the electronic, but not the nuclear, states change;
\item	and finally, including the full extent of off-diagonal terms in $\mH_{\R{non-diag}}$ (\zfr{fig5}D), gives rise to anti-crossings at each of the former crossing points, although with differing energy gaps (described in \zsr{cascade}E and App. A).
\eenum

To label eigenstates in \zr{eq5}, we assume that $\mS$ labels the $2^N$ possible nuclear spin states in each electronic manifold consisting of all combinations of spin up or down. For instance, for $N{=}2$, we have $\mS{=} \{\ket{\dw\dw},\ket{\dw\up}, \ket{\up\dw},\ket{\up\up}\}$.  These also correspond to the eigenstates of the diagonal Hamiltonian $\mH_{\R{diag}}$. Using Hamming binary notation with ${h_j}{=}$ 0 or 1 representing $\ket{\up}$ or $\ket{\dw}$ respectively, the eigenenergies of the diagonal Hamiltonian in the $m_s{=}\{0,\R{+}1\}$ manifolds take the forms,
\beq
E_J^{(0)}(\xo_{\R{MW}}) {=} \sum_{j=1}^{N}(-1)^{h_j}\xo_j^{(0)}\\
\zl{eq7}
\eeq
\vspace{-6mm}
\beq
E_J^{(1)}(\xo_{\R{MW}}) {=} (\xD + \xg_e B_0 - \xo_{\R{MW}}) + \sum_{j=1}^{N}(-1)^{h_j}\xo_j^{(1)}\: .
\zl{eq8}
\eeq
This allows us to constrain the locations of the LZ-LACs in the checkboard. The index $J$ here runs over $2^N$ and we note from \zr{eq7} that the eigenvalues are independent of $\xo_{\R{MW}}$ in the $m_s{=}0$ manifold, and scale linearly with it in the $m_s{=}$+1 case (\zr{eq8}). While the discussion here is applicable for an arbitrary hyperfine network $A_j^{\pll}$, in this paper we consider an exemplary scenario where hyperfine couplings are in a \I{non-degenerate} configuration such that,
\beq
\xo_j^{(0)}=j^{p}\:;\:\xo_j^{(1)}={\alpha}j^{p}.
\zl{eq9}
\eeq 
For example, in \zfr{fig5}, we employ this model with ${\alpha}{=}5$ and the exponent $p{=}1.1$.  The lack of degeneracies in this model illuminate the physics of LZ traversals more clearly, while the same general principles follow in the degenerate case as well.

 \vspace{-2mm}
\subsection{Checkerboard of anti-crossings $\mI_{k,\ell}$}
 \vspace{-1mm}
Let us now return to the situation in \zfr{fig5}B and determine the exact positions of the LZ-LACs. Eigenenergies in each of the $m_s{=}\{0,\R{+}1\}$ manifolds form separate sets of parallel lines, and lines corresponding to different manifolds intersect. For $N$ nuclear spins connected to the central NV center, there are in general $2^{2N}$ crossing points. Together, these crossing points form the checkerboard pattern  $\mI_{k,\ell}$ (as in \zfr{fig1}D and \zfr{fig4}) in the 2D plot of energy and frequency.

Let $\mI$ refer to the matrix of coordinates of these crossing points in a diagram as in \zfr{fig4}, where the \I{abscissa} $\mI_{k,\ell}(1)$ and \I{ordinate} $\mI_{k,\ell}(2)$ refer to frequency and energy respectively. The label indices $k,\ell \in 1{\cdots}2^{N}$ refer to levels in the $m_s{=}{\R{+}1}$ and $m_s{=}0$ manifolds respectively. We arrange the crossings from left-to-right and top-to-bottom (i.e. increasing in frequency and descending in energy). Referring to \zfr{fig4}, the element $\mI_{1,1}$, for instance, refers to the crossing between levels $\ket{0,\up\up\up\up}$ and $\ket{\R{+}1,\dw\dw\dw\dw}$. Following this convention, the elements can be written as,
\beq
\mI_{k,\ell} = \lb E_{k}^{(1)}(0)- E_{\ell}^{(0)}, E_{\ell}^{(0)}\rb
\zl{eq10}
\eeq 
given that from \zr{eq7},
\beq
E_k^{(1)}(\xo_{\R{MW}}) = E_k^{(1)}(0)- \xo_{\R{MW}}
\zl{eq11}
\eeq
and thus, at the LZ-LACs,
\beq
\xo_{\R{MW}} = E_{k}^{(1)}(0)- E_{\ell}^{(0)}.
\zl{eq12}
\eeq

 For instance, the LZ-LAC $\mI_{2,1}{=} \lb E_{2}^{(1)} (0)- E_1^{(0)}, E_1^{(0)} \rb$ is formed from the intersection of two levels in the $m_s{=}\{0,\R{+}1\}$ manifolds. The $\mI_{k,\ell}$ checkerboard pattern is displayed in \zfr{fig1}D and \zfr{fig4}, and is \I{tilted} since the crossings corresponding to lower nuclear states occur at higher frequency, $\mI_{k,{\ell}+1}(1){>}\mI_{k,\ell}(1)$. A key consequence is that the system, upon a MW sweep, encounters LZ-LACs in a \I{sequential} manner (\zfr{fig1}D) — important for ultimately causing \I{``bias"} in the nuclear polarization buildup.

 \vspace{-2mm}
\subsection{LACs: Energy gaps, symmetries and hierarchies}
 \vspace{-1mm}
Continuing our discussion onto \zfr{fig5}C-D, we now elucidate the emergence of the LZ-LACs and their associated energy gaps.  We will refer to the checkerboard energy gaps at positions (${k, \ell}$) as $\vxe_{k,\ell}$, using the same notation convention as above. The gaps arise approximately at the abscissa locations marked by $\mI_{k,\ell}(1)$. Numerically extracted gap center locations are plotted as points in \zfr{fig1}D and \zfr{fig4}, while the dashed lines refer to the checkerboard crossing points obtained using an approach similar to \zfr{fig5}C. For simplicity, we refer to the $\mI_{k,k}$ positions on the checkerboard as those on the \I{diagonal}, while those at the positions $\mI_{k,2^N-k+1}$ belong to the \I{“conjugate diagonal”} (see \zfr{fig4}). 

To intuitively unravel the formation of the LZ-LACs, consider first the action of the Rabi frequency term $\xO_e$, as in \zfr{fig5}C. Energy gaps open up only at the conjugate diagonal (for positive $A_{j}^{\pll}$), corresponding to the anti-crossings between states $\ket{0,\mS}{\rt}\ket{\R{+}1,\mS}$, where $\mS$ is any state of the nuclear spins. The energy gaps here $\vxe_{k,2^N-k+1}{=}\xO_e$, since this transition just corresponds to a single electron flip — akin to the situation encountered in a rapid adiabatic passage over the electrons. Including the perpendicular hyperfine terms $A_{j}^{\pp}$ in \zr{eq6} provides additional LZ-LACs at each of the $\mI_{k,\ell}$ crossing points in \zfr{fig5}D. 

App. A shows a detailed evaluation of the corresponding gaps, but we comment here that they  satisfy two \I{mirror symmetry} conditions (evident in \zfr{fig4}), 
\bea
|\mO - \mI_{k,\ell}|&=& |\mO - \mI_{2^N-k+1,2^N-\ell+1}| \zl{eq13}\\
\vxe_{k,\ell} &=& \vxe_{2^N-k+1,2^N-\ell+1}\zl{eq14}
\eea
\benum[label=(\roman*)]	
		\item The first condition (\zr{eq13}) describes a mirror symmetry of the checkerboard diagram with respect to the center $\mO = (\xD+\xg_e B,0)$ (see \zfr{fig4}), a consequence of the spin-1/2  $\Cs$ nuclei and the fact that all $A_{j}^{\pll}$ are positive in the chosen model in \zr{eq9}.  
	\item Similarly, \zr{eq14} elucidates a symmetry in the energy gaps about the conjugate diagonal on the checkerboard (see \zfr{fig4}). For instance, the gaps corresponding to LZ-LACs $\ket{0,\up\up\up}{\ltrt}\ket{\R{+}1,\dw\dw\up}$ and $\ket{0,\dw\dw\up}{\ltrt}\ket{\R{+}1,\up\up\up}$ should be identical because they just differ in an exchange between nuclear labels undergoing the crossing.
	\eenum

\begin{figure}[t]
  \centering
 \includegraphics[width=0.49\textwidth]{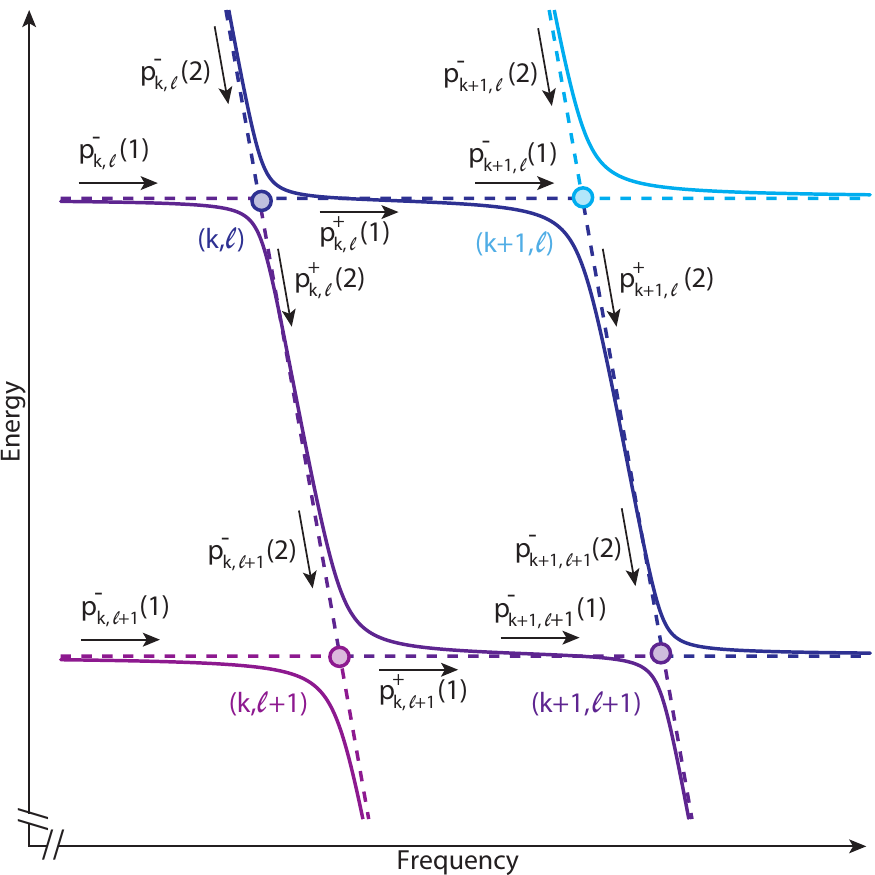}
  \caption{\T{Transfer matrix formalism. } Panel shows a schematic of traversal through a Galton system of anti-crossings.  We consider a representative section of a larger $\mI_{k,\ell}$ checkerboard as in \zfr{fig4}, but focus here on only four LZ points (dots). Populations bifurcate at the LZ points similar to a Galton board (\zfr{fig3}) upon a driven low-high frequency MW sweep. Components of $\T{p}_{k,\ell}^{\mp}$, two-component column vectors denoting nuclear populations entering (or leaving) the node $(k,\ell)$, are shown. Bifurcation of the populations is described by the transfer matrix $\T{T}_{k,\ell}$ (\zr{eq19}). We assume there is no system evolution between successive LZ points; it follows that $p_{k,\ell}^{+}(1) {=}p_{k+1,\ell}^{-}(1)$ and $p_{k,\ell}^{+}(2) {=}p_{k,\ell+1}^{-}(2)$, and yields a recursion relationship connecting nearest-neighbor populations (\zr{eq21}). This permits a simple evaluation of system evolution in terms of nuclear population redistribution probabilities upon traversal through the Galton board.} 
\zfl{fig6}
\end{figure}

 \vspace{-2mm}
\section{Transfer matrix formalism}
\zsl{transfer}
 \vspace{-2mm}
\subsection{Traversals through cascaded Landau-Zener LACs}
 \vspace{-1mm}
We now develop the analogy with the Galton board (see \zfr{fig3}) to evaluate the evolution of nuclear populations through the checkerboard $\mI_{k,\ell}$, and show how this evolution yields nuclear hyperpolarization. We make two operational assumptions: 
\benum[label=(\roman*)]
\item The $\mI_{k,\ell}$ LZ-LAC points are assumed to be hit sequentially and their effects are evaluated individually. This is reasonable if the energy gaps $\vxe_{k,\ell}$ are small compared to their frequency separation. 
\item NV (re)polarization is assumed to happen far away from the exact LZ-LAC points $\mI_{k,\ell}$, a good approximation when $\vxe_{k,\ell}{<}\mB$. This allows us to consider the LZ system traversals independent of optical pumping.
\eenum
At each checkerboard point, let $\T{p}_{k,\ell}^{-}$ and $\T{p}_{k,\ell}^{+}$ denote the nuclear populations of the two states undergoing the LZ-LAC, \I{before} and \I{after} the anti-crossing is encountered. This is shown in \zfr{fig6} for four LZ crossings in a 2x2 portion of the $\mI_{k,\ell}$ checkerboard. We denote the populations as column vectors, with the first (second) element referring to nuclear populations in the $m_s{=}0$ ($\R{+}1$) manifolds respectively,
\beq
\T{p}_{k,\ell}^{-}=\left[\begin{array}{c}
p_{k, \ell}^{-}(1) \\
p_{k, \ell}^{-}(2) \\
\end{array}\right]\:;\: 
\T{p}_{k,\ell}^{+}=\left[\begin{array}{c}
p_{k, \ell}^{+}(1) \\
p_{k, \ell}^{+}(2) \\
\end{array}\right] 
\zl{eq15} .
\eeq
The effect of any chosen trajectory through (as in \zfr{fig4}) the LZ cascade can then be built out of simple 2x2 matrices. Evolution starts with the population restricted to the $m_s{=}0$ manifold, and with the $\Cs$ nuclear spins in a mixed state $\T{1}_N$. This means that each of the nuclear states in the $m_s{=}0$ manifold have an equal initial starting probability. Traversal can then be calculated by considering populations starting with the initial states,
$
\T{p}_{1,\ell}^{-} = \left[\begin{array}{c}
1 \\
0 \\
\end{array}\right]
$
and averaging the result over the $2^N$ $\ell$ states. 

 \vspace{-2mm}
\subsection{Transfer matrix formalism}
 \vspace{-1mm}
Consider first the action of a single LZ-LAC point $\mI_{k,\ell}$ with energy gap $\vxe_{k,\ell}$. Encountering it leads to a rearrangement of nuclear populations,
\bea
p_{k,\ell}^{+}(1)&=&\eta_{k,\ell} p_{k,\ell}^{-}(1)+\left(1-\eta_{k,\ell}\right) p_{k,\ell}^{-}(2)\non \\
p_{k,\ell}^{+}(2)&=&\left(1-\eta_{k,\ell}\right) p_{k,\ell}^{-}(1)+\eta_{k,\ell} p_{k,\ell}^{-}(2)
\zl{eq16}
\eea
where $\eta_{k,\ell}$ refers to the tunneling probability through the anti-crossing and has the form,
\beq
\eta_{k,\ell}{=}\exp(-\vxe_{k,\ell}^2/\dxo_{\R{MW}})
\zl{eq17} .
\eeq 
Here $\dxo_{\R{MW}}$ is the MW sweep rate, and $\dxo_{\R{MW}}{=}\mB{\xo}_r$ for the linear chirp employed (see \zfr{fig2}A).
\zr{eq16} can then be recast in terms of a transfer matrix $\T{T}_{k,\ell}$,
\beq
 \T{p}_{k,\ell}^{+}=\T{T}_{k,\ell}\: \T{p}_{k,\ell}^{-}\:,
\zl{eq18}
\eeq
where, 
\beq
\T{T}_{k,\ell} =\mat{\eta_{k,\ell}}{\left(1- \eta_{k,\ell}\right)}{\left(1- \eta_{k,\ell}\right)}{\eta_{k,\ell}}\:.
\zl{eq19}
\eeq
$\T{T}_{k,\ell}$ plays the role of a \I{``redistribution''} operator since it causes the rearrangement of populations at the $\mI_{k,\ell}$ anti-crossing points (\zfr{fig6}). Traversal is \I{adiabatic} if $\dxo_{\R{MW}}\vxe_{k,\ell}^{-2}{\ll} 1$ and \I{diabatic} if $\dxo_{\R{MW}}\vxe_{k,\ell}^{-2}{\gtrsim}1$. In the limiting case when the gap is large ($\eta_{k,\ell}{\rt} 0$) and there is a complete transfer of population between the states at the LZ-LAC, $\T{T}_{k,\ell} = \mat{0}{1}{1}{0}$. On the other hand, for a purely diabatic transfer, $\T{T}_{k,\ell} =\T{1}_{2}=\mat{1}{0}{0}{1}$, and the populations are preserved. In a more generic, yet diabatic case, there is a bifurcation of populations at the $\mI_{k,\ell}$ crossing point (\zfr{fig6}).
Note that the matrix formulation above only considers the redistribution of nuclear populations at each LZ-LAC and ignores coherences. In reality, there is a phase picked up by the quantum state as it traverses through an LZ-LAC. However, in our experiments (to a good approximation), this phase can be ignored because:
\benum[label=(\roman*)]
\item 	In the high defect density sample we consider in \zfr{fig2}, the electronic $T_{2e}^{*}{\sim}50{\R{ns}}{\ll}\xo_r^{-1}$ is short~\cite{Acosta09} - and much smaller than the single sweep time $\xo_r^{-1}$ as well as the traversal time between successive $\mI_{k,\ell}$ points. 
\item 	The bare nuclear coherence time over which electrons are repolarized — $T_{2n}^{*}{\sim}$1ms — is of the order of the sweep period, but much smaller than the total polarization time (which involves ${>}10^3$ sweeps) (\zfr{fig2}A). To a good approximation then, any nuclear coherences produced as a result of the traversal rapidly die away due to internuclear interactions.
\item 	In experiments (\zfr{fig2}C), we perform an ensemble average over a large collection of $e$-$n$ systems like in \zfr{fig1}A. In this case, the coherence of the electronic state can be considered randomized~\cite{Wenckebach17} and one can consider only the redistributions of state populations at each LZ-LAC.
\eenum

\zr{eq18}-\zr{eq19} make more concrete the analogy to the Galton board (\zfr{fig3}).  A \I{classical} Galton board operates through balls falling through a system of pegs under gravity (see \zfr{fig3}A). At each peg, the balls can bounce left or right (usually with 50\% probability each). In a similar manner, the LZ-LACs here form the ``pegs'', the ``balls'' are the nuclear populations, and the swept MWs provide the driving force analogous to gravity (\zfr{fig3}B). The LZ Galton board here is {tilted}, and the nuclear populations bounce at each LZ-LAC either right or down, the probability being conditioned on the size of the corresponding energy gap $\vxe_{k,\ell}$.

Building on this analogy,  let us now evaluate the action of sequentially encountering LZ anti-crossings under a \I{low-to-high} frequency MW sweep. Extending \zr{eq18}, the net traversal can then be written as a product of transfer matrices corresponding to a trajectory through the cascaded LZ-LACs. Since there is no population evolution between successive LZ-LACs,
$
p_{k,\ell}^{-}(1){=}p_{k-1, \ell}^{+}(1)$ and $p_{k,\ell}^{-}(2){=}p_{k, \ell-1}^{+}(2).
$
Therefore \zr{eq18} can be rewritten as,
\beq
\T{p}_{k,\ell}^{+}=\T{T}_{k,\ell}\: \T{p}_{k,\ell}^{-} = \T{T}_{k,\ell}\left[\begin{array}{c}
p_{k-1, \ell}^{+}(1) \\
p_{k, \ell-1}^{+}(2) \\
\end{array}\right]
\zl{eq20}
\eeq
or equivalently,
\beq
\T{p}_{k,\ell}^{+}=\T{T}_{k,\ell} \T{M}_{r} \T{p}_{k-1,\ell}^{+}+\T{T}_{k,\ell} \T{M}_{d} \T{p}_{k,\ell-1}^{+}
\zl{eq21}
\eeq
where we define the \I{``walk''} operators,
\beq
\T{M}_{d}=\mat{0}{0}{0}{1}\:;\:\T{M}_{r}=\mat{1}{0}{0}{0}.
\zl{eq22}
\eeq
$\T{M}_{d}$ and $\T{M}_{r}$ refer to walks over the checkerboard by a single-step \I{``down"} or \I{``right"} respectively. \zr{eq21} is interesting because it connects populations at LZ-LAC sites $(k,\ell)$ with prior nearest neighbor sites on the checkerboard, schematically described in \zfr{fig6}. Combining \zr{eq18} and \zr{eq21} provides the \I{recursive} relationship,
\bea
\T{p}_{k,\ell}^{+}&=&\T{T}_{k,\ell}\T{M}_{r} \T{T}_{k-1,\ell}\T{M}_{r}\T{p}_{k-2,\ell}^{+} + \T{T}_{k,\ell}\T{M}_{r} \T{T}_{k-1,\ell}\T{M}_{d}\T{p}_{k-1,\ell-1}^{+}\non\\
&+&\T{T}_{k,\ell}\T{M}_{d} \T{T}_{k,\ell-1}\T{M}_{r}\T{p}_{k-1,\ell-1}^{+} + \T{T}_{k,\ell}\T{M}_{d} \T{T}_{k,\ell-1}\T{M}_{d}\T{p}_{k-1,\ell-2}^{+}\non ,
\eea
which connects an $\mI_{k,\ell}$ checkerboard point to its nearest and next-nearest neighbors (see \zfr{fig6}). This process can be continued to encompass every path of evolution through the cascaded LZ system. Consider for instance the portion of the trajectory $\T{\I{v}}$ (black line) in \zfr{fig4}B. The final nuclear population through it is then, $\T{p}_{4,3}^{+} = \T{T}_{\R{eff}}\T{p}_{1,1}^{-}$, where $\T{p}_{1,1}^{-}=\left[\begin{array}{c}
1 \\
0 \\
\end{array}\right]$ where,
\beq
\T{T}_{\R{eff}}= \T{M}_{r} \T{T}_{43} \T{M}_{r} \T{T}_{33} \T{M}_{r} \T{T}_{23} \T{M}_{d} \T{T}_{22}\T{M}_{d} \T{T}_{21} \T{M}_{r} \T{T}_{11} \T{M}_{r} .
\zl{23}
\eeq
In general, consider a path consisting of a sequence of nearest neighbor points through $\mI_{k,\ell}$: $v{=}\{v_j\} {=} \{(k_1,\ell_1),\cdots,(k_j,\ell_j)\cdots (k_{n},\ell_{n})\}$. If $\T{p}_i^{-}$ and $\T{p}_f^{+}$ are the populations at the initial and final coordinates respectively, this recursive condition can be generalized,
\bea
\T{p}_f^{+} &=& \T{T}_{v_n}\T{M}_{(v_{n},v_{n-1})}\cdots\T{T}_{v_3}\T{M}_{(v_3,v_{2})}\T{T}_{v_2}\T{M}_{(v_2,v_{1})}\T{T}_{v_1}\T{M}_{r}\cdot \T{p}_i^{-}\non\\
 &=& \prod_{j=1}^{n} \T{T}_{v_j}\T{M}_{(v_{j+1},v_{j})}\cdot \T{p}_i^{-}
\zl{eq24}
\eea
where $\T{T}_{v_j}$ is the corresponding bounce operator at coordinate $v_j{=}(k_j,\ell_j)$, and the operator $\T{M}$ refers to the walk through the checkerboard,
\beq
\T{M}_{(v_{j+1},v_{j})}=\left\{
                \begin{array}{ll}
                  \T{M}_r& \text{for $k_{j+1} = k_j+1$}\\
                  \T{M}_d& \text{for $\ell_{j+1} = \ell_j+1$}\\
                \end{array}
              \right.
\zl{eq25} 
\eeq
based on whether the subsequent trajectory entails a movement right or down in the board. From \zr{eq21} and \zr{eq22}, and since $ \T{p}_i=\left[\begin{array}{c}
1 \\
0 \\
\end{array}\right]$, it is now possible to evaluate the final probability of the path defined by the coordinates $v=\{v_j\}$ as,
\beq
\mP=\prod_{j=1}^{n}\mC_{v_{j}}
\zl{eq26} ,
\eeq
with coefficients,
\beq
\mC_{v_{j}}=\left\{
                \begin{array}{ll}
                  \eta_{k_j,\ell_j}& \text{if $k_{j-1} = k_{j+1}$ or $\ell_{j-1} = \ell_{j+1}$}\\
                  (1-\eta_{k_j,\ell_j})& \text{if $k_{j-1} \neq k_{j+1}$ and $\ell_{j-1} \neq \ell_{j+1}$}\\
                \end{array}
              \right.
\zl{eq27} .
\eeq
These two factors entering \zr{eq27} can be intuitively understood. If at a specific LZ-LAC the trajectory $\T{\I{v}}$ involves a path that continues straight and does not take a \I{“bend”}, then the term that enters the product is the direct tunneling probability $\eta_{k,\ell}$. Alternatively, a bend appears with the factor $(1-\eta_{k,\ell})$. This allows for the final probability $\mP$ (and thus final population) for any path to be easily found.  For simplicity, we decompose $\mP$ at any LZ-LAC point into two components, $\mP=[\mP^{(1)}, \mP^{(2)}]^T$, with $\mP^{(1)}$ representing final $m_s{=}$0 populations and $\mP^{(2)}$ representing final $m_s{=}$\R{+}1 populations (see \zfr{fig7}). For instance, when applied to the full trajectory $\T{\I{v}}$ in \zfr{fig4} (exiting in the $m_s{=}$+1 manifold), one can just write down the trajectory,
\bea
\mP^{(2)}{=}\eta_{1,1} (1-\eta_{2,1}) \eta_{2,2} (1-\eta_{2,3}) 
\eta_{3,3} (1-\eta_{4,3}) \eta_{4,4} \eta_{4,5} \eta_{4,6} 
\eta_{4,7} \non \\
\eta_{4,8} \eta_{4,9} \eta_{4,10} (1-\eta_{4,11}) 
\eta_{5,11} (1-\eta_{6,11}) \eta_{6,12} \eta_{6,13} \eta_{6,14} 
\eta_{6,15} \eta_{6,16} \non .
\eea
Taking identical $\eta_{k,\ell}$ values for every energy gap $\vxe_{k,\ell}$ gives the simple expression,
\bea
\mP^{(2)}{=}{\eta}^{16} (1-\eta)^5 \non .
\eea
This tractable method of tracking nuclear spin populations is the key result of this paper.

\begin{figure}[t]
  \centering
 \includegraphics[width=0.49\textwidth]{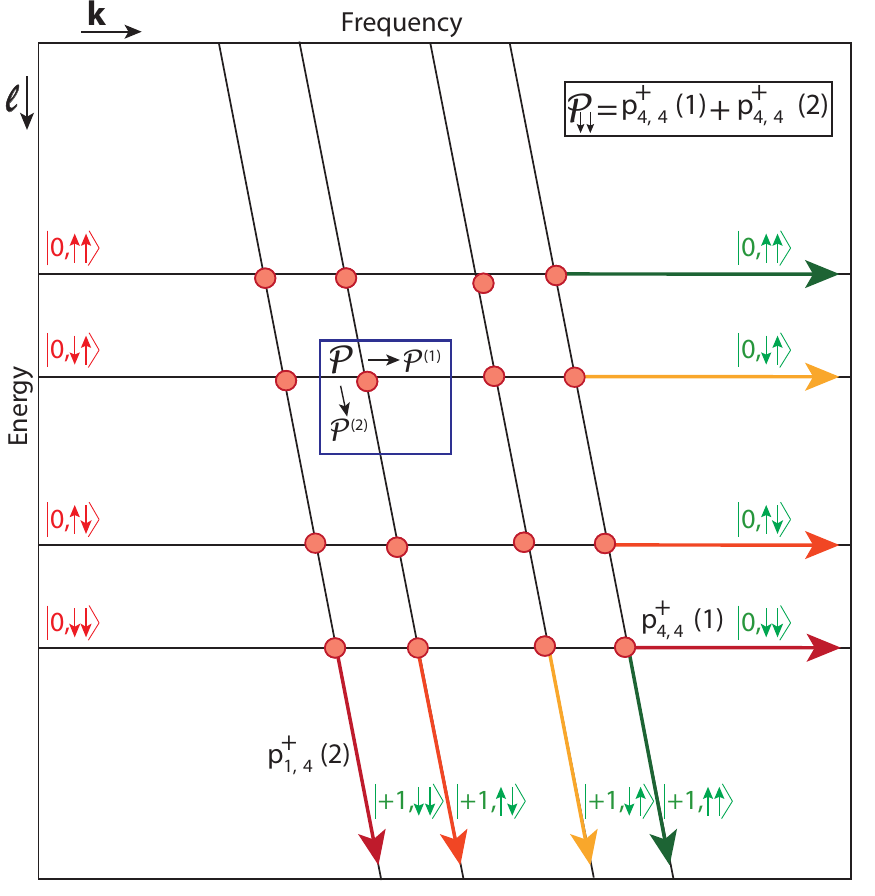}
  \caption{\T{Calculation schematic for transfer probability $\mP$.} Representative LZ Galton board showing a schematic of population evolution for $N{=}2$ nuclei. Energy levels are marked. Panel shows the probability $\mP$ of arrival at an LZ-LAC (\I{boxed}), and probabilities $\mP^{(1)}$ and $\mP^{(2)}$ of bifurcation \I{right} or \I{down} (traveling towards the $m_s{=}0$ and $m_s{=}$+1 manifolds respectively).  Colored arrows represent contributions to the same nuclear state $\mS$ from different electronic manifolds, which are eventually combined due to the action of the laser (\zr{eq29}).  For example,  the probability of  ending in the $\ket{\dw\dw}$ state (assuming unit value starting population) is $\mP_{\dw\dw}=p^{+}_{4,4}(1) +  p^{+}_{4,4}(2)$.  Contributing populations from both electronic manifolds,  $p^{+}_{4,4}(1)$ and $p^{+}_{1,4}(2)$,  are highlighted by red arrows in this case. }
\zfl{fig7}
\end{figure}

 \vspace{-2mm}
\subsection{General traversal through the $\mI_{k,\ell}$ checkerboard}
 \vspace{-1mm}
 Following from \zr{eq24}-\zr{eq27}, given a final coordinate in the $\mI_{k,\ell}$ checkerboard, it is possible to retrospectively determine all paths that lead to it, and hence the transfer probability as the sum over all the paths the system can traverse. Considering traversal from an initial coordinate $(k_i, \ell_i)$ to a final coordinate $(k_f,\ell_f)$, there are in general $(\ell_f{-}\ell_i)$ vertical  steps and $(k_f{-}k_i)$ horizontal steps required. This gives rise to $L=(k_f+\ell_f - k_i-\ell_i)$ total traversal steps, and there are hence $L_p{=}\binom{L}{k_f-k_i}$ total paths of the traversal, where $\lb \cdot\rb$ is the binomial operator. Following \zr{eq26}, we can write the traversal probability from $(k_i, \ell_i)$ to $(k_f, \ell_f)$ as,
\beq
\mP[(k_i,\ell_i){\rt} (k_f,\ell_f)] {=} \sum_{\{v\}\in L_p}\lb \prod_{j=1}^{n}\mC_{v_{j}}\rb .
\zl{eq28}
\eeq
\zr{eq28} generalizes finding traversal probabilities for any trajectory over vertices $\T{\I{v}}$, where the index $j$ represents the position on the path.

 \vspace{-2mm}
\subsection{Electron repolarization: Action of laser}
 \vspace{-1mm}
Key to the hyperpolarization process is the ability to regenerate the populations of the NV electrons by means of optical pumping.  This permits the ability to repeatedly accumulate polarization in the nuclear spins, using them to indirectly report on the ESR spectrum of the electrons. We assume the action of the laser instantaneously resets the population of the $m_s{=\R{+}1}$ NV state to $m_s{=}0$, while keeping nuclear populations unaffected (e.g. $\ket{\R{+}1,\mS}{\rt}\ket{0,\mS}$). If $\T{p}_{k, \ell}^{'}$ denotes nuclear populations after the action of the laser, then for \I{all} points $(k,\ell)$ on the LZ-LAC checkerboard, $
\T{p}_{k, \ell}^{'\pm}{=}\T{L}\cdot \T{p}_{k, \ell}^{\pm}
$, with laser action described by the operator, 
\beq
\T{L} {=} \mat{1}{1}{0}{0}\:.
\zl{eq29}
\eeq
We are making the assumption here that the extent of each of the LZ anti-crossing regions (defined by the energy gaps $\vxe_{k,\ell}$) on the checkerboard is small in comparison to the region between them. Note also that in the DNP protocol in \zfr{fig2}A, the laser is applied continuously with the MW sweep. In principle, therefore, the action of the laser can occur at any point during the traversal through the LZ checkerboard. The essential physics are, however, unaffected if we consider the laser reset as just occurring at the end (or beginning) of the traversal through the entire LZ checkerboard. This physically means that finding the net populations after each cycle requires us to evaluate the sum over corresponding populations in both electronic manifolds. This is schematically shown in \zfr{fig7}, where for instance, the population of nuclei reaching the state $\ket{\dw\dw}$ is calculated by the sum,  $\mP_{\dw\dw}=p^{+}_{4,4}(1) +  p^{+}_{4,4}(2)$.

 \vspace{-2mm}
\section{Galton board hyperpolarization}
\zsl{Galton}
 \vspace{-2mm}
 
\subsection{Calculating the nuclear polarization $P$}
 \vspace{-1mm}
Elucidating the magnitude of polarization is essential to understanding the origin of nuclear polarization \I{``bias"}, and is key to this is calculating traversal probability through the LZ Galton board. Using \zr{eq28}, we can find the population at any final point $(k_f,\ell_f)$, which then permits finding the population of any nuclear state $\mS$ after traversal through an LZ cascade. Since each nuclear state appears once for \I{each} electronic manifold, the imbalance upon system traversal can be accounted for by evaluating the total population in each state at any time instant through the diagram. The population in the $m_s{=}0$ state then takes the form (see \zr{eq16}),
\beq
{p}_{2^{N},\ell_{f}}^{+}(1) {=} \sum_{n=1}^{2^{N}} \mP^{(1)}[(1,n){\rt} (2^{N},\ell_f)] ,
\zl{eq30}
\eeq
where the sum indicates the total probability starting from the left column of the diagram. Similarly,  the population in the $m_s{=}$+1 state, assuming $A^{\pll}{>}0$, takes the form,
\beq
{p}_{k_f,2^{N}}^{+}(2) {=} \sum_{n=1}^{2^{N}} \mP^{(2)}[(1,n){\rt} (k_f,2^{N})] .
\zl{eq31}
\eeq
Due to the action of the laser (\zr{eq29}), these populations are effectively merged and the population of any nuclear state $\mS$ is,
\beq
\mP_{n} {=} {p}_{2^{N},\ell_{n}}^{+}(1) + {p}_{k_n,2^{N}}^{+}(2)
\zl{eq32}
\eeq
where the subscript $n \in 1{\cdots}2^{N}$ here indexes the state $\mS$, and the individual terms can be calculated following \zr{eq28}. For instance, this is schematically shown for the state $\mP_{\dw\dw}$ in \zfr{fig7}. Finally, the nuclear polarization $P$ is defined as the net \I{excess} of population in the nuclear down states compared to the up states, and for the case of $A^{\pll}{>}0$ that we consider,
\beq
P{=}\sum_{n=1}^{2^{N-1}}{\mP}_n - \sum_{n=2^{N-1}+1}^{2^{N}}{\mP}_n 
\zl{eq33} .
\eeq

 \vspace{-2mm}
\subsection{Special case: Single-spin ratchet}
 \vspace{-1mm}
\zsl{ratchet}
As a simple application, consider an NV center coupled to a single $\Cs$ nuclear spin. The mechanism here was referred to as a \I{“spin ratchet”} in Ref.~\cite{Ajoy17}. We note that while this case ($N{=}1$) was considered before in Ref.~\cite{Zangara18}, the Galton board analogies in \zfr{fig8} provide a simple means to analyze the bifurcation and recombinations of the populations in a intuitive and graphical manner. The analogy also provides a means to generalize the mechanism of operation to larger $N$, an apriori non-trivial problem that has not been previously considered elsewhere. 

In \zfr{fig8},  we consider both cases of hyperfine coupling $A^{\pll}{>}0$ (\zfr{fig8}A) and $A^{\pll}{<}0$ (\zfr{fig8}B), with \zfr{fig8}C-D showing the corresponding $\mI_{k,\ell}$ checkerboards. Apart from the reversed order of levels in the $m_s{=}$+1 manifold, the structures in these diagrams are similar. This inversion arises from the difference in order in which the LZ-LACs corresponding to the ``small" and ``large" energy gaps are encountered during a MW sweep.

\begin{figure}[t]
  \centering
 \includegraphics[width=0.49\textwidth]{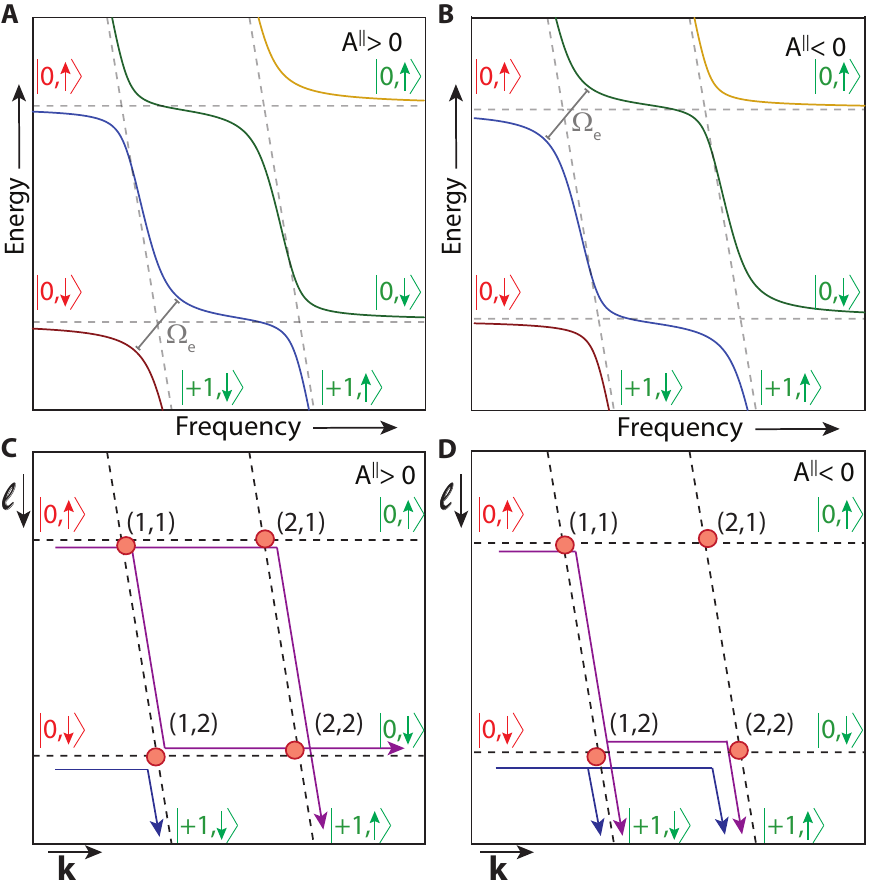}
  \caption{\T{\I{``Spin ratchet"} hyperpolarization} corresponding to the case of an NV center coupled to a \I{single} $\Cs$ nuclear spin $(N{=}1)$. We consider two cases, (A) $A^{\pll}{>}0$ and (B) $A^{\pll}{<}0$. Panels show LZ anti-crossings similar to \zfr{fig1}D, from which we abstract the checkerboards in the lower panels.  Nuclear states are shown in either case.  Large energy gaps corresponding to $\xO_e$ are highlighted. (C-D) \I{LZ checkerboards} corresponding to panels (A) and (B) respectively which elucidate the change in the order of energy levels in the $m_s{=}$+1 manifold, and the resulting change in the order in which the small and large energy gaps are encountered. We assume here that large energy gaps have $\eta{=}0$ and thus do not allow direct tunneling. The possible trajectories of populations starting in the $\ket{0,\up}$ and $\ket{0,\dw}$ states are shown (purple and blue arrows respectively).  Polarization buildup occurs only in the former case ($A^{\pll}{>}0$), while interference of the traversals through the Galton system leads to a destruction of nuclear polarization buildup in the latter case ($A^{\pll}{<}0$). }
\zfl{fig8}
\end{figure}

\noindent\T{\I{(i)} Case of $A^{\pll}{>}0$:} Consider a single traversal from left-to-right starting from populations restricted to the $m_s{=}0$ manifold and ending with NV repolarization. Using \zr{eq32}, and following from \zr{eq28}, the probability of the nuclear state $\ket{\dw}$ remaining in $\ket{\dw}$ can be written as,
\beq
\mP(\dw\: \rightarrow\: \dw) = {p}_{2,2}^{+}(1) + {p}_{1,2}^{+}(2) , \non
\eeq
taking,
\beq
{p}_{2,2}^{+}(1) = \mP^{(1)}[(1,2){\rt} (2,2)], \:\R{and}\:,
{p}_{1,2}^{+}(2) = \mP^{(2)}[(1,2)] , \non
\eeq
in the Galton  board in \zfr{fig8}C. This provides us a method to calculate the individual probabilities as,
\bea
\mP(\dw\: \rightarrow\: \dw)&=&\left(1-\eta_{1,2}\right)+\eta_{1,2} \eta_{1,1} \non\\
\mP(\dw\: \rightarrow\: \up)&=&\eta_{1,2}\left(1-\eta_{1,1}\right) \non\\
\mP(\up\: \rightarrow\: \dw)&=&\eta_{1,2}\left(1-\eta_{1,1}\right)+2 \eta_{1,1}\left(1-\eta_{1,1}\right)\left(1-\eta_{1,2}\right) \non\\
\mP(\up\:\rightarrow\: \up)&=&\eta_{1,1}\eta_{1,2}+\eta_{1,1}^{2}\left(1-\eta_{1,2}\right)+\left(1-\eta_{1,1}\right)^{2}\left(1-\eta_{1,2}\right) , \non
\eea
where the tunneling probabilities follow \zr{eq17} and we have exploited symmetries in energy gaps (described in \zsr{cascade}E). Each term in these expressions corresponds to a \I{different} trajectory through $\mI_{k,\ell}$. For instance, there are two paths that constitute the term, $\mP(\dw\: \rightarrow\: \dw)$, corresponding to the probabilities of $\ket{0,\dw}{\rt}\ket{0,\dw}$ and $\ket{0,\dw}{\rt}\ket{\R{+}1,\dw}$. This permits a means to determine the nuclear hyperpolarization, where we evaluate the difference in populations between the nuclear states at the end of the sweep as, 
\bea
P&=&\lsb \mP(\dw\: \rightarrow\: \dw) + \mP(\up\: \rightarrow\: \dw)\rsb -\lsb \mP(\dw\:\rightarrow\: \up) + \mP(\up\:\rightarrow\: \up)\rsb\non\\
&=&\left(1-\eta_{1,2}\right)\left[1-\left(2 \eta_{1,1}-1\right)^{2}\right]
\zl{eq34} .
\eea
Ultimately, the expression in \zr{eq34} demonstrates that the polarization levels are obtained as the difference of the probabilities of transitions to spin down states and transitions to spin up states. 

It is simplest to evaluate \zr{eq34} for the situation when the Rabi frequency $\xO_e$ is large so that the traversals through the energy gaps $\vxe_{1,2}$ and $\vxe_{2,1}$ are adiabatic, but those through the others are diabatic. We then have $P{=}1-\left(2 \eta_{1,1}-1\right)^{2}$; a non-zero hyperpolarization develops through an accumulation of nuclear spins in the $\ket{\dw}$ state with a single MW sweep. Building on \zr{eq34} then, the net hyperpolarization developed in a total time $T$ (\zfr{fig2}A) is,
\beq
P_{\R{net}}=\left[1-\exp \left(\frac{-w_{L}}{\omega_{r}}\right)\right] \cdot T \omega_{r} \cdot\left(1-\eta_{1,2}\right)\left[1-\left(2 \eta_{1,1}-1\right)^{2}\right],
\zl{eq35}
\eeq
where the first term encapsulates the starting electron polarization, the second term ($\propto T\xo_r$) gives the total sweeps in time $T$, and the last term is \zr{eq34}. The mechanism can therefore be thought of as a  \I{“ratchet”} — every MW sweep develops a finite amount of polarization.

\noindent \T{\I{(ii)} Case of $A^{\pll}{<}0$:} A similar analysis can be carried out for the situation with $A^{\pll}{<}0$ (see \zfr{fig8}B), yielding,
\bea
\mP(\dw\: \rightarrow\: \dw)&=&\eta_{1,1}\eta_{1,2} + \eta_{1,2} (1-\eta_{1,1}) \non\\
\mP(\dw\: \rightarrow\: \up)&=&\left(1-\eta_{1,2}\right) \non\\
\mP(\up\: \rightarrow\: \dw)&=&\left(1-\eta_{1,1}\right)^2(1-\eta_{1,2})+{\eta_{1,1}}^2(1-\eta_{1,2})\non\\
&+&2\eta_{1,1}(1-\eta_{1,1})(1-\eta_{1,2})\non\\
\mP(\up\: \rightarrow\: \up)&=&\eta_{1,1}\eta_{1,2} + \eta_{1,2}(1-\eta_{1,1})\non,
\eea
once again exploiting symmetrical energy gaps.
The polarization developed upon one MW sweep (similar to \zr{eq34}) now gives,
\bea
P&=&\lsb \mP(\dw\: \rightarrow\: \dw) + \mP(\up\: \rightarrow\: \dw)\rsb -\lsb \mP(\dw\:\rightarrow\: \up) + \mP(\up\:\rightarrow\: \up)\rsb\non\\
&=&0 .
\zl{eq36}
\eea
There is therefore no net buildup of polarization in this case, a consequence of a fine balance between the population traversals in the two arms of the Galton board in \zfr{fig8}B.

\noindent \T{\I{(iii)} Sweep direction dependence:} The discussion above is centered on low-to-high frequency sweeps for the two cases of positive and negative hyperfine coupling. Consider now the opposite scenario to \zfr{fig8}A-B - where sweeps are applied from high-to-low frequency. The order of the two energy gaps in \T{\I{(i)}} and \T{\I{(ii)}} above are now reversed. This results in \T{\I{(i)}} no effective hyperpolarization in the $A^{\pll}{>}0$ case, and \T{\I{(ii)}} hyperpolarization in the $\ket{\up}$ state in the $A^{\pll}{<}0$ case. The expressions are identical to that in \zr{eq34}. We note that this results in two consequences that are borne out by experiments (see companion paper~\cite{Pillai21}):
\benum[noitemsep,topsep=0pt,label=(\roman*)]
	\item	 Due to equal proportions of $A^{\pll}{<}0$ and $A^{\pll}{>}0$ present in experiments, there is an inversion in the hyperpolarization sign upon a reversal in the direction of the MW sweep. This is in contrast with other DNP methods (e.g solid effect) where the DNP sign is different for the two halves of the spectrum. 
	\item A consequence is that all parts of the spectrum yield the same sign in the $\Cs$ hyperpolarization signal, making unravelling the underlying electronic spectrum a simpler experimental undertaking.
	\item Finally,  in the case of large $N$ systems,  one would expect a slight shift in the frequency center of the DNP derived spectrum in the case of one sweep direction with respect to the other. 
\eenum

\begin{figure*}[t]
  \centering
 \includegraphics[width=1\textwidth]{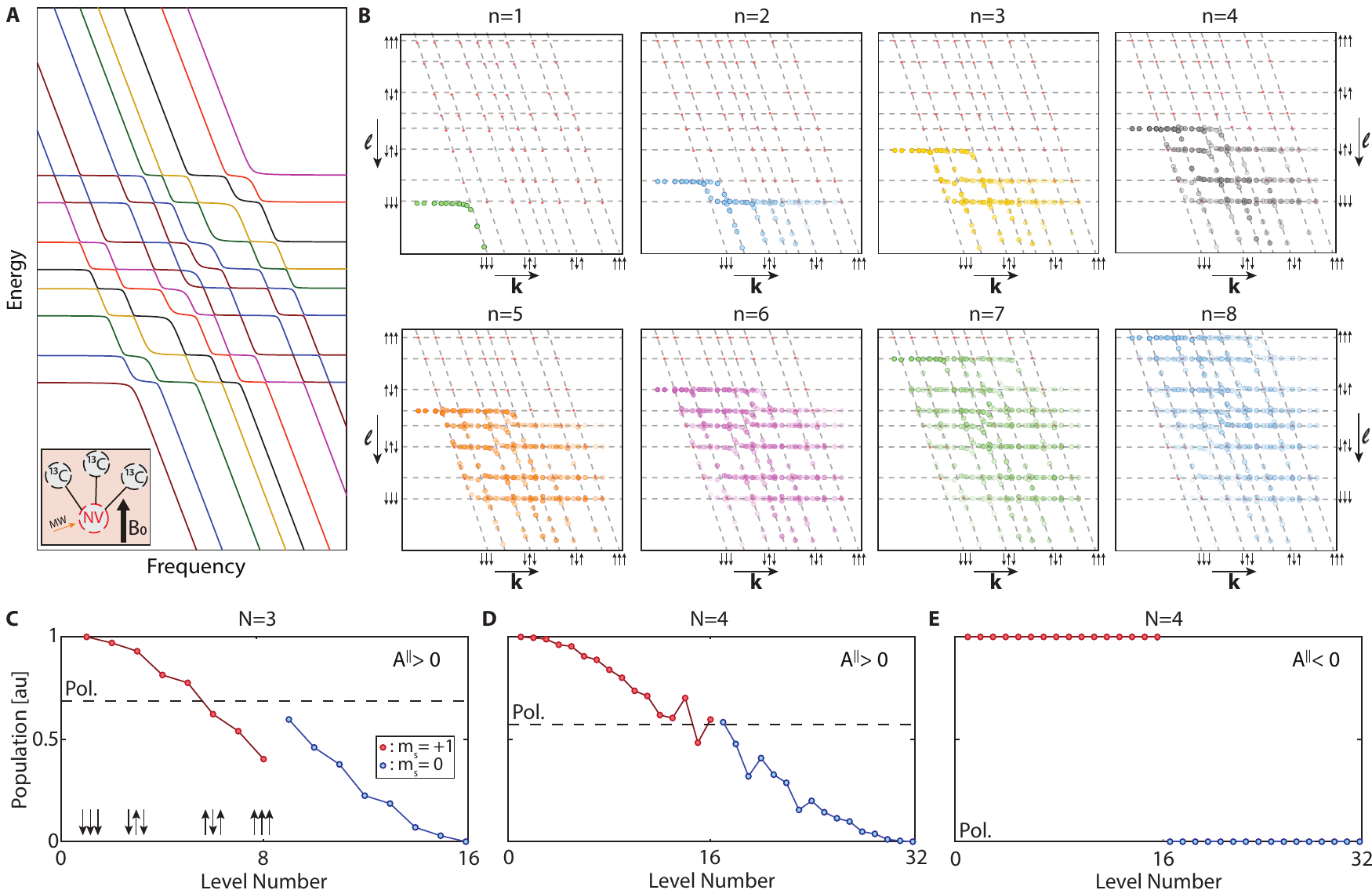}
  \caption{\T{Evolution of nuclear populations through the LZ Galton board}.  Panels show the emergence of nuclear hyperpolarization upon a low-to-high frequency MW sweep (assuming $\eta_{k,2^{N}-k+1}{=}0$ along the conjugate diagonal). (A) \I{LZ cascade} for $N{=}3$ nuclei (shown in inset) and $A^{\pll}{>}0$ as in \zfr{fig5}D. (B) \I{Traversal through LZ cascade.} We extract the LZ-LAC points (red) in (A) to define the LZ Galton board (dashed lines). Representative nuclear states are marked. Panels show evolution of nuclear populations (colored points), starting from different nuclear states $\mS$ (indexed by $n$ in a Hamming notation) in the $m_s{=}0$ manifold, under a MW sweep. Opacity of dots represents relative population. There is a clear bias towards nuclear down states. (C) \I{Population bias and emergence of hyperpolarization}. Final populations in different nuclear states (numbered in a Hamming ordering) in the $m_s{=}0$ and $m_s{=}$+1 manifolds (blue and red points respectively) after the traversals through the LZ Galton board shown in (B). Dashed line represents the hyperpolarization level.  (D) Panel shows populations similar to (C), but for $N{=}4$. (E) Panel shows populations similar to (D), but now for $A^{\pll}{<}0$. Due to the change in the order of energy levels and energy gaps, there is no net polarization developed (see also \zfr{fig8}D). }
\zfl{fig9}
\end{figure*}

 \vspace{-2mm}
\subsection{Origin of nuclear polarization bias}
 \vspace{-1mm}
The origin of hyperpolarization can be traced to the \I{differential} traversal through the $\mI_{k,\ell}$ checkerboard in a manner that \I{“biases”} the system to increase population in particular nuclear states at the cost of others. Considering the example of an NV coupled to a single nuclear spin, \zr{eq34} shows a biased traversal that follows,
\beq
p_{f}(\ket{\dw}) \app p_{i}(\ket{\dw}) + \beta p_{i}(\ket{\up}) \:;\: p_{f}(\ket{\up}) \app (1-\beta)p_{i}(\ket{\up}),
\zl{eq37}
\eeq
where $\xb$ is a finite probability and $p_{i}(\cdot)$ and $p_{f}(\cdot)$ refer to nuclear populations before and after traversal through the LZ Galton board, and have the form, 
\bea
p_{i}(\ket{\dw}) &=& p_{1,2}^{-}(1); 
p_{i}(\ket{\up}) = p_{1,1}^{-}(1);\non \\
p_{f}(\ket{\dw}) &=& p_{2,2}^{+}(1) + p_{1,2}^{+}(2);
p_{f}(\ket{\up}) = p_{2,1}^{+}(1) + p_{2,2}^{+}(2)\non ,
\eea
Effectively, one of the nuclear states (here $\ket{\dw}$) is unaffected, while the other (here $\ket{\up}$) is flipped with the probability $\beta$. Successive sweeps through the LZ system then set up a ``\I{ratchet}’’ that builds nuclear hyperpolarization in the $\ket{\dw}$ state. 

To understand the origin of this bias for larger $N$, consider traversal through the full checkerboard in \zfr{fig4}. As we demonstrate in \zfr{fig9}, a combined consequence of \I{(i)} the large energy gaps along the conjugate diagonal, \I{(ii)} the order that the energy gaps are encountered in, and \I{(iii)} the tilted nature of the LZ Galton board, there is a natural bias in evolution towards the left and bottom of the diagram (\zfr{fig9}). This net excess of population in the nuclear down states over the nuclear up states (\zr{eq33}) ultimately results in the  hyperpolarization that we measure in experiments (e.g. in \zfr{fig2}C and Ref.~\cite{Pillai21}). In \zsr{Galton}E, we illustrate the development of this bias via numerical simulations of LZ-LAC traversals (\zfr{fig9}). 

 \vspace{-2mm}
\subsection{Analogy to Mach-Zender interferometry}
 \vspace{-1mm}
 We note finally that the Galton board illustrated in the panels in \zfr{fig8}C-D bear resemblance to the construction of a Mach-Zender (MZ) interferometer~\cite{Zetie00}, where the LZ-LACs serve analogous to beam splitters. One might then consider the fine cancellation occurring in \zfr{fig8}D as an interference effect between the two arms of the MZ interferometer. A more detailed discussion of this connection is beyond the scope of this manuscript; however, we note that such connections could elevate the LZ-LAC checkerboard, naturally occurring in spin systems as in \zfr{fig1} to applications exploiting the power of parallelized interferences in cascaded MZ interferometers (e.g. BosonSampling~\cite{Aaronson04}).

 \vspace{-2mm}
\subsection{Numerical simulations}
 \vspace{-1mm}
The formalism developed via \zr{eq26},  \zr{eq28}, \zr{eq32}, and \zr{eq33} permits a tractable solution for the traversals through a LZ-LAC Galton board. Before an analytical solution, we first consider a numerical evaluation in \zfr{fig9}, considering the case of $N{=}3$ and $A^{\pll}{>}0$ - and assuming for simplicity $\xt_j{=}\pi/2$ - with the LZ cascade structure shown in \zfr{fig9}A. We consider evolution under a low-to-high frequency MW sweep, and solve the traversal following \zr{eq32}. We assume that, as in experiments, the Rabi frequency $\xO_e$ is large, such that the energy gaps at the conjugate diagonals are large and traversals through them are adiabatic. The individual panels in \zfr{fig9}B then show the populations of nuclear states at each point on the $\mI_{k,\ell}$ checkerboard, starting with populations confined to starting $m_s{=}0$ nuclear states $\mS$ (indexed by $n$). We track the populations (dots) after each of the $\mI_{k,\ell}$ LZ-LACs (red points), where the opacity of the dots denotes the \I{normalized} population level. The analogy to the classical Galton board is evident here (\zfr{fig3}A); there is a \I{``sieving"} effect as the populations bifurcate at each LZ-LAC, yielding progressively smaller populations as one traverses deeper into the checkerboard. There is an intrinsic bias towards the \I{left} and \I{bottom} of the Galton board (evident in \zfr{fig9}B); this ultimately yields hyperpolarization. 

\zfr{fig9}C makes this more clear by plotting the resulting nuclear state populations in the $m_s{=}0$ and $m_s{=}$+1 manifolds (blue and red points respectively) following a sweep over the \I{full} Galton board.  \zfr{fig9}D shows an analogous calculation for $N{=}4$.  Representative nuclear states are marked and state $n$ follows a Hamming ordering. In both panels, a decrease in population is evident with increasing state number $n$ in each manifold, while the population of a particular nuclear state is higher in the $m_s{=}$+1 (red) manifold than in the $m_s{=}0$ (blue) one. Both these observations reflect the bias in the Galton board traversal. Ultimately, upon application of the laser to repolarize all the electronic polarization to $m_s{=}0$, one obtains nuclear hyperpolarization, the value of which is denoted by the black dashed lines in \zfr{fig9}C-D.

Finally, \zfr{fig9}F considers the inverted scenario where $A^{\pll}{<}0$ (for an identical Galton board) and a full MW sweep is carried out. In this case, while the LZ cascade has a similar structure to the diagram in \zfr{fig9}A, the order of the nuclear states in the $m_s{=}$+1 manifold are reversed, and the large energy gaps appear on the diagonal instead of the conjugate diagonal. As a result, the development of polarization bias is impeded through a fine counterbalance of the nuclear populations as they traverse the Galton board. A type of destructive interference then ensues, resulting in no hyperpolarization in this case. This is shown in \zfr{fig9}E, where we consider the populations in each of the nuclear states similar to \zfr{fig9}C-D, assuming the diagonal energy gaps are large so that traversals through them are adiabatic. All of the nuclear states in each manifold then have the same population, and there is no net development of hyperpolarization. We note, however, that the situations for $A^{\pll}{>}0$ and $A^{\pll}{<}0$ are exactly reversed when the MW sweep is applied from high-to-low frequency. 

\begin{figure}[t]
  \centering
 \includegraphics[width=0.49\textwidth]{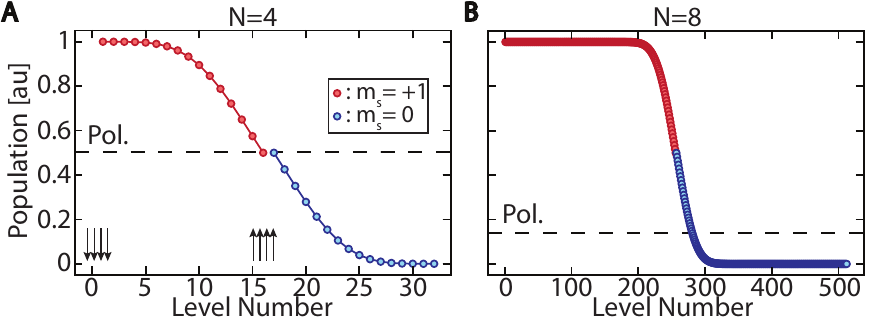}
  \caption{\T{Analytic solution to LZ Galton board traversal.} Panels show the evaluation of the analytic expression \zr{eq39} for the solution of total nuclear populations following a full MW sweep through a (A) $N{=}4$ and (B) $N{=}8$ LZ Galton board. Similar to \zfr{fig9}, the populations ($2^{N} \mP_{n}$) of nuclear states in the $m_s{=}0$ and $m_s{=}$+1 manifolds (blue and red points respectively) are shown, with $\mP_n$ following \zr{eq39}. We have assumed here $p{=}q{=}0.5$ (see \zfr{fig3}B). Dashed lines represent the polarization level. Panels demonstrate close resemblance to the numerical simulations in \zfr{fig9}.}
\zfl{fig10}
\end{figure}

 \vspace{-2mm}
\subsection{Theoretical evaluation of Galton board traversal}
 \vspace{-1mm}
The numerical solution to traversals through an LZ Galton board in \zfr{fig9}, while revealing the essential physics of the biased traversal and buildup of hyperpolarization, is still unwieldy to solve for large $N$.  This is because of the exponentially scaling number of LZ-LACs the system entails ($\propto 2^{2N})$.  That said however, the analogy to the classical Galton board developed above (see \zfr{fig3}),  makes feasible a full \I{analytic} solution under certain simplifying assumptions.  Consider first that the initial nuclear populations are equally distributed among states in the $m_s{=}0$ electronic manifold before entering the LZ Galton board at points $(1,\ell)$. Nuclear populations travel along horizontal lines $\ell$ and vertical lines $k$ in the LZ Galton board and redistribute at every LZ-LAC on $\mI_{k, \ell}$, where $k, \ell \in 1 {\cdots} 2^{N}$. Similar to the assumptions made in a classical Galton board,  we will define the probability of redistribution down and right to be $p$ and $q$ respectively at every LZ-LAC, with $p{+}q{=}1$. In addition, for points along the conjugate diagonal, we assume that the energy gaps are large enough that $\eta_{k,2^{N}-k+1}{=}0$, and thus traversals are adiabatic. We demonstrate below that these assumptions permit us to capture the physics of the cascaded LZ-LAC traversal while permitting an analytical solution (\zfr{fig10}).

Towards this end,  we evaluate the Galton board traversal to determine the probability of the total nuclear population that ends up in each specific spin state $\mS$, indexed by $n$ (where $n \in 1 {\cdots} 2^{N}$). Due to the action of the laser (\zr{eq29}), the nuclear population in any $n$ state is then just the sum of the population that exits point $(n,2^{N})$ downward and point $(2^{N},2^{N}{-}n{+}1)$ to the right (as described in \zr{eq32}).

We first consider the traversal of a nuclear population starting at a point (1,$\ell$) before cascading through the board. The population entering at point $(1,2^{N})$ will necessarily travel down, due to the large energy gap it immediately encounters. We can express this with a function $f(n)$ which is equal to 1 when $n{=}1$ and is equal to 0 otherwise.

For nuclear populations starting at any other $\ell \in 1 {\cdots} 2^{N}{-}1$, we can now employ logic similar to that in a classical Galton board to evaluate the probabilities of each spin state. For states ending in the $m_s{=}$+1 manifold, any nuclear population must travel $2^{N}{-}\ell{-}1$ steps down and $n{-}2$ steps right (disregarding movement through the conjugate diagonal). Because the nuclear population can take a variety of trajectories (e.g. \zfr{fig4}) to reach the same spin state, we can use a combinatoric to find the number of ways to choose $n{-}2$ right steps from the total steps. Thus, we can write the probability as the exact form, ${(n-2)+(2^{N}-\ell-1)\choose n-2}p^{2^{N}-\ell}q^{n-2}$,  where $\cdot\choose{\cdot}$ is a combinatoric operator,  revealing a binomial form similar to a classical Galton board. Likewise, we can compute the probability for reaching a spin state $n$ on the right (in the $m_s{=}0$ manifold), which requires $2^{N}-n-\ell$ down steps and $2^{N}-2$ right steps. This probability is written as ${(2^{N}-n-\ell)+(2^{N}-2)\choose 2^{N}-n-\ell}p^{2^{N}-n-\ell}q^{2^{N}-1}$.  Combining both these results,  we can write the total probability that a nuclear population from a point $(1,\ell)$, where $\ell \in 1 {\cdots} 2^{N}{-}1$, will end up in a specific spin state $n$ as,

\bea
\mP_{n,\ell}={(n{-}2){+}(2^{N}{-}\ell{-}1)\choose n{-}2}p^{2^{N}{-}\ell}q^{n{-}2} \non\\
{+}{(2^{N}{-}n{-}\ell)+(2^{N}{-}2)\choose 2^{N}{-}n{-}\ell}p^{2^{N}{-}n{-}\ell}q^{2^{N}{-}1}\zl{eq38} ,
\eea

where $\mP_{n,\ell} {=} \mP^{(1)}[(1,\ell){\rt} (2^{N},n)] + \mP^{(2)}[(1,\ell){\rt} (n,2^{N})]$ as in \zr{eq38}. The total proportion of nuclear population that ends up in a specific state $n$ is then just the average of the probabilities from the above formula for all possible entrance points, and takes the form,
\begin{widetext}
\bea
\mP_{n}=\frac{1}{2^{N}}\Bigg(\sum_{\ell=1}^{2^{N}-1}\Bigg[{(n{-}2){+}(2^{N}{-}\ell{-}1)\choose n{-}2}p^{2^{N}{-}\ell}q^{n{-}2}{+}{(2^{N}{-}n{-}\ell)+(2^{N}{-}2)\choose 2^{N}{-}n{-}\ell}p^{2^{N}{-}n{-}\ell}q^{2^{N}{-}1}\Bigg]+f(n)\Bigg)\zl{eq39} ,
\eea
\end{widetext}
where $\mP_{n}$ is the same probability as in \zr{eq32}. \zfr{fig10} displays the result of this analytic expression (\zr{eq39}) for a LZ Galton board of size $N{=}4$ and $N{=}8$. We emphasize the strong resemblance to the numerical simulations in \zfr{fig9}C-D.  This also illustrates that the essential physics of the LZ Galton traversal and the development of hyperpolarization is less affected by the exact energy gaps at the individual LZ-LACs.

Interesting to note is the case of the reverse sweep (high-to-low frequency), in which the population starts from the right side of the Galton board and traverses through the board by moving either up or left at each LZ-LAC with probabilities $p$ and $q$. Because this is the same board and all other above assumptions apply, it becomes clear that there is symmetry between the two cases. Specifically, the reverse sweep is identical in behavior to the original sweep when the board is \I{flipped} across the conjugate diagonal. This means that we can use the same analytical solutions we have acquired for the original sweep to the reverse sweep, with a relabeling of the states as $s$ according to the symmetrical flip (where $s{=}2^N{-}n$). This takes the form,

\begin{widetext}
\bea
P_{s}{=}\frac{1}{2^{N}}\Bigg(\sum_{\ell=1}^{2^{N}-1}\Bigg[{(2^{N}{-}s{-}2){+}(2^{N}{-}\ell{-}1)\choose 2^{N}{-}s{-}2}p^{2^{N}{-}\ell}q^{2^{N}{-}s{-}2} {+}{(s{-}\ell)+(2^{N}{-}2)\choose s{-}\ell}p^{s{-}\ell}q^{2^{N}{-}1}\Bigg]+f(2^{N}{-}s)\Bigg)\zl{eq40} .
\eea
\end{widetext}
An evaluation of \zr{eq40} then shows, by symmetry,  the development of opposite polarization in this case, matching experimental results~\cite{Pillai21}.

\begin{figure*}[t]
  \centering
 \includegraphics[width=1\textwidth]{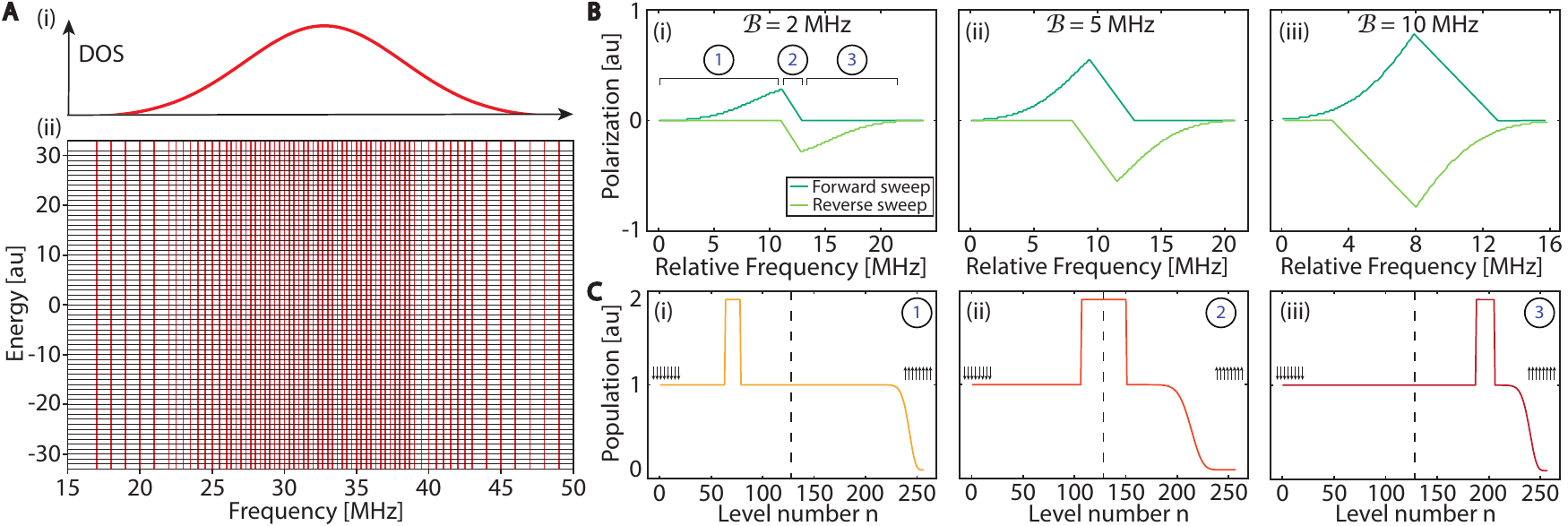}
  \caption{\T{Numerical simulation showing \I{``ESR-via-NMR"} spectral mapping}. (A) Lower panel shows an exemplary LZ Galton board with $N{=}8$ nuclei employed in simulations. Energy levels in the $m_s{=}$+1 manifold have a Gaussian DOS of width ${\app}$13.5MHz \I{(upper panel)} similar to \zfr{fig2}B. Units chosen here for the Galton board  extent are arbitrary and just used for illustration. Since numerical evaluation is performed column wise, the tilt of the Galton board is not shown here. (B) \I{Numerical result} showing the developed $\Cs$ hyperpolarization using sweep windows of size $\mB$ of increasing size through the Galton board in (A), similar to the experiments in \zfr{fig2}C. Forward (dark) and reverse (light) sweeps result in positive and negative polarization respectively. Increasing $\mB$ yields a greater hyperpolarization signal and a broader spectral profile, matching experimental observations (\zfr{fig2}C and Ref.~\cite{Pillai21}).  For narrow $\mB$ (panel \I{(i)}) we find three distinct regimes in the hyperpolarization profiles (\I{circled}), where there is a sharp increase, a rapid linear decrease and no net polarization respectively. With increasing $\mB$, the profiles begin to approximate the DOS profile of the underlying Galton board being swept over. (C) \I{Unraveling the origin} of the hyperpolarization profile in panel B(i). Representative nuclear states are marked.  For each of the three regimes, we plot the resulting nuclear populations in the $2^N{=}$256 nuclear states, numbered $n$ following a Hamming convention. Hyperpolarization is evaluated as the difference in populations between two halves of this plot (indicated by the dashed line), following \zr{eq33}. In regime 1 \I{(i)}, there is an emergence of hyperpolarization because the population imbalances constructively add. In regime 3 \I{(iii)}, however, the population imbalances are in the same half of the plot, yielding no net polarization.}
\zfl{fig11}
\end{figure*}

 \vspace{-2mm}
\subsection{Evaluation of a window sweep on the Galton board}
 \vspace{-1mm}
While the previous section explores the dynamics of a \I{full} sweep of the Galton board, we now turn our focus to the case when a small window $\mB$ is swept (corresponding to experiments in \zfr{fig2}C).  Analytic solutions similar to \zr{eq28} are still tractable, but due to expressions being considerably more complex than the full sweep behavior,  here we instead carry out a numerical solution. We employ the same assumptions above that each LZ-LAC is encountered sequentially and that the energy gaps along the conjugate diagonal are large. First, motivated by experiment, we assume a Gaussian distribution for energy levels (DOS) in the $m_s{=}$+1 manifold, as seen in \zfr{fig11}A. This modulates the spacing of the $\mI_{k, \ell}$ columns and impacts the \I{number} of $m_s{=}$+1 energy levels that are swept in any specific frequency window $\mB$.

We employ a physically motivated method of evaluating traversals through the board. First note that the ability to uncover spectral DOS mapping dynamics necessarily requires simulations of large $N$ Galton boards to prevent artifacts stemming from edge effects in small $N$ systems. For large Galton boards as in \zfr{fig11}A (which is for $N{=}8$ nuclei, thus having 256 nuclear states) however, the full simulation approach employed in \zfr{fig9} proves expensive and unwieldy due to the exponential scaling system size. Instead here, for simplicity, we choose a specific order in which the LZ-LACs are evaluated — solving the result of traversals through each $\mI_{k, \ell}$ column from the top to the bottom, working through the board from left to right. This is an acceptable assumption because the bifurcation action at any LZ-LAC is only dependent on previously encountered LZ-LACs above or to the left of it. With this additional assumption in place, we can evaluate the traversals of populations, starting and ending within a specified window $\mB$ defined by its starting column and width. We choose a large Galton board (N=8) in \zfr{fig11}A, and since evaluations are column wise here, we show the board without its tilt.

Following this approach, \zfr{fig11}B shows the numerical evaluations of the traversal of the board in \zfr{fig11}A with varying $\mB$ and both forward (dark) and reverse (light) sweeps. These plots are produced by taking repeated window sweeps of a set size $\mB$ while gradually moving the starting point (corresponding to the starting point $f_0$ in \zfr{fig2}) through the board and calculating the polarization after the population in each state has redistributed following \zr{eq28}. With increasing window size (e.g. in \zfr{fig11}B(ii)-(iii)), we observe that the obtained hyperpolarization profiles approximately trace the underlying electronic DOS of the underlying LZ Galton board, in this case the Gaussian function chosen. There is also an increase in overall polarization as the window size $\mB$ increases, and an increasing width in the hyperpolarization profile with increasing $\mB$ (see \zfr{fig11}B). Both these features match experimental observations (\zfr{fig2}C and Ref.~\cite{Pillai21}). 

Why does the observed hyperpolarization signal approximately track the electronic spectral DOS? Insight into this is revealed in \zfr{fig11}B(i) and \zfr{fig11}C. For narrow window sizes $\mB$, the simulations in \zfr{fig11}B(i) reveal that one half of the Galton board produces positive nuclear polarization, while the other half produces no net polarization. Moving from left-to-right in \zfr{fig11}B(i) for instance, we identify three regimes in the obtained hyperpolarization profiles: a regime \circled{1}, in which there is a increase in polarization; regime \circled{2}, in which there is a linear decrease; and regime \circled{3} with zero net polarization. As the window size $\mB$ increases, the differences in the profiles between the three regimes become less prominent, and ultimately in \zfr{fig11}C(iii) the profile appears Gaussian-like. At this point, the  profiles approximate the underlying spectral DOS in the starting LZ Galton board.  

To unravel the internal dynamics of the Galton traversals that yield \zfr{fig11}B, let us start by considering a representative sweep in regime \circled{1} of \zfr{fig11}B(i). The result is shown in \zfr{fig11}C(i), where we plot the nuclear populations in each of the $2^N$ levels (labeled $n$) after the sweep over $\mB$ is completed. In this regime, it is predominantly the \I{net} nuclear-down states that are being swept over. A (positive) increase in the nuclear populations is then evident in nuclear states that are being swept over; this arises from the populations effectively moving towards the bottom of the LZ-Galton board. Similarly, there is a decrease of populations in high-$n$ nuclear states, corresponding to predominantly nuclear up states. Since the final polarization plotted in \zfr{fig11}B(i) is evaluated (following \zr{eq33}) as the \I{difference} in populations between the nuclear states on either half of the plot in \zfr{fig11}C(i) (denoted by the dashed line), there is a net development of hyperpolarization in this case. Indeed, \zfr{fig11}C(i) suggests that the polarization produced is approximately proportional to the number of levels swept over, since the “positive” portion of this population imbalance has a width that scales with $\mB$.   

In a similar manner, \zfr{fig11}C(ii) shows a sweep over a window $\mB$ that contains a mix of net nuclear-down and up states (regime \circled{2}). In this case, the positive and negative population regions move closer together (see \zfr{fig11}C(ii)).  A difference (across the dashed line) then yields the approximately linear decrease in polarization observed in regime \circled{2} as the sweep window passes into the net nuclear-up states.  Finally, an illuminating view is provided by a sweep carried out in regime \circled{3}), shown in \zfr{fig11}C(iii). Here, both population imbalances occur in the \I{same} half of the plot; hence there is no hyperpolarization buildup. Ultimately a combination of these features yields the profile in \zfr{fig11}B(i). It is illustrative that the reverse sweep (light line in \zfr{fig11}B(i)) has a mirror-inverted profile about the Galton board center. Finally, upon increasing $\mB$ (\zfr{fig11}B(ii)-(iii)), the sharp differences between the three regimes in \zfr{fig11}B(i) become less prominent; since the profile becomes a convolution with the (larger) sweep window. In this case, both positive and negative sweeps yield profiles that approximately map the spectral DOS.

 \vspace{-2mm}
\section{Outlook}
 \vspace{-1mm}
Our work opens many interesting directions for future research. First, the formalism developed here allows one to model the effect of a driven $N$-spin quantum system undergoing evolution through a cascaded series of level anti-crossings. Given that the number of anti-crossings scales exponentially with system size $\propto 2^{2N}$, this problem appears intractable at first. However, the description in terms of traversals of an analogous \I{``Galton board”} permits both analytic and numeric solutions, allowing one to extract physically relevant information in the large $N$ limit. The cascaded LZ-LAC structure studied here (\zfr{fig1}) is a motif that occurs commonly in several contexts, such as in quantum walks in Hilbert spaces~\cite{Baum85,Boutis04,Ajoy12b,Burgarth17} and operator scrambling~\cite{Krojanski04,Swingle16}, as well as in applications such as BosonSampling~\cite{Aaronson11} that involve multiple cascaded Mach-Zender interferometer traversals~\cite{Tillmann13,Spring13,Wang17} (see \zfr{fig8}). We therefore envision that the approach developed here might contribute means to employ engineered central spin systems, such as in \zfr{fig1}, to study these phenomena. 

More direct applications of this work lie in mechanistic descriptions of dynamic nuclear polarization (DNP) in the large nuclear spin limit: the predominant experimental regime of interest. Our work shows a systematic approach to extend DNP models beyond standard descriptions of $e$-$n$ or $e$-$e$-$n$ systems and can inform physics at large $N$~\cite{Wenckebach17}. Apart from the MW driven DNP problem considered here, we envision applications to hyperpolarization mechanisms employing field sweeps~\cite{Henshaw19}, or with magic angle spinning~\cite{Mentink15,Thurber12,Mentink12}, where a similar cascaded structure of LACs is encountered~\cite{Pravdivtsev14,Pravdivtsev15}. Similar descriptions to DNP arise in the context of quantum state \I{``cooling”} in a variety of systems~\cite{Schliesser08,Hamann98,Teufel11}; we therefore envision wider applications of this methodology developed here in these systems.

An application we considered was in performing \I{“ESR-via-NMR”}~\cite{Pillai21}, wherein the electronic spectrum is effectively coaxed out by only reading out the nuclear populations. This relies on mapping the electronic spectral density of states (DOS) to nuclear population imbalances via hyperpolarization transfer (similar to \zfr{fig2}C). The utility of this approach may be particularly relevant in systems where ESR spectra may be inaccessible, e.g when the electrons are so dilute that ESR signals from them are too weak, or their short $T_{2e}$ makes probing them directly more difficult. Instead, one could leverage the more abundant, longer-lived (high $T_{1n}$) nuclear spins in their immediate surroundings as a means to indirectly probe them by accumulating electronic DOS information in nuclear polarization prior to readout. We envision particularly compelling applications in photo-polarizable electronic systems in  molecular complexes~\cite{Rugg19}.

From a technological perspective, the approach developed here portends applications in quantum memories~\cite{Bradley19} constructed out of central spin systems (such as \zfr{fig1}A), wherein information from the electronic spins is mapped in a \I{“one-to-many”} fashion to long-lived nuclear registers. Similarly, there are interesting applications in quantum sensing, including modalities where the electronic spin sensors are interrogated via polarization transfer of surrounding nuclear spins. In the context of NV centers, this opens applications in RF interrogated magnetometry, wherein the NV center populations are readout by $\Cs$ nuclei without a MW cavity (see accompanying manuscript~\cite{Pillai21}).  As opposed to conventional optically interrogated NV magnetometers, this can permit DC magnetometers that function in turbid or optically dense media, harnessing the immunity of RF readout to optical scattering, permitting the arbitrary orientation of the diamond crystals. This will open avenues for a networked quantum sensing \I{“underwater”}~\cite{Pillai21}, with a variety of applications (e.g. undersea magnetic anomaly detection~\cite{Ge20}).

 \vspace{-3mm}
\section{Conclusion}
 \vspace{-1mm}
In conclusion, we theoretically considered the problem of polarization transfer in a system of an electronic spin coupled to $N$ surrounding nuclei via frequency swept microwaves. We described how, in an appropriate rotating frame, the system can be described by traversals through a cascade of Landau-Zener anti-crossings (LZ-LACs). We showed that the traversals can be mapped into an equivalent evolution through a \I{“Galton board”}, where individual LZ-LACs map onto ``pegs”, upon encountering which the nuclear populations bifurcate; this makes the solution of the complex dynamics through a coupled $e$-$n$ system consisting of $2^{2N}$ anti-crossings analytically tractable. Our formal solution procedure relied on elucidating trajectories of evolution through the LZ Galton board via transfer matrices, allowing one to extract insights into the development of \I{``bias’’} in the traversals that then leads to hyperpolarization. An implication is that spin bath polarization levels track the electronic spectral DOS (\zfr{fig2}C), allowing the nuclear spins to indirectly report on the underlying electronic spectrum, in effect achieving \I{``ESR-via-NMR”}~\cite{Pillai21}. The Galton board mapping developed here suggests that a similar approach can be extended to problems involving traversals through a series of cascaded energy crossings, with applications in DNP~\cite{Pravdivtsev14}, spin qubit or resonator “cooling”~\cite{Schliesser08,Teufel11} and in quantum sensing and information processing with central-spin systems.

 \vspace{-5mm}
\subsection*{Acknowledgments}
 \vspace{-1mm}
We gratefully acknowledge discussions with J. Reimer, C. Meriles and P. Zangara. This work was funded by ONR under N00014-20-1-2806.

\appendix

\begin{figure}[t]
  \centering
 \includegraphics[width=0.35\textwidth]{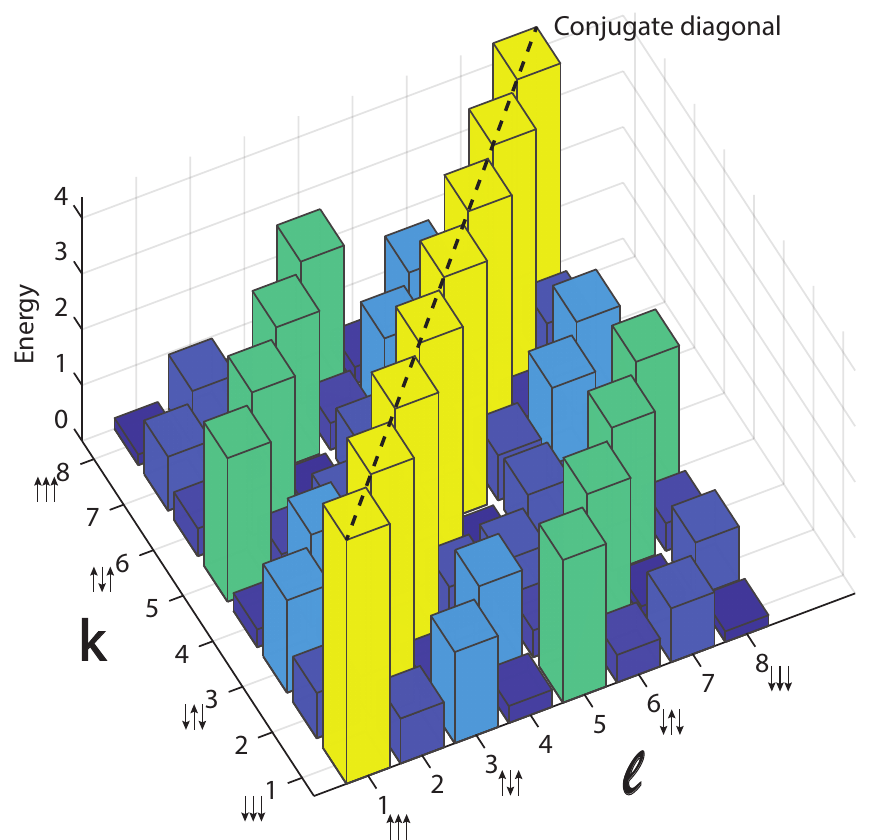}
  \caption{\T{Energy of the LZ anti-crossings $\vxe_{k,\ell}$} here plotted with respect to their positions on the $\mI_{k,\ell}$ checkerboard for an $N{=}3$ system.  The conjugate diagonal is illustrated by the dashed line and representative nuclear states are marked.  We focus here on the scaling of the energy gaps $\vxe_{k,\ell}$ based on the number of nuclear spin flips required, hence we fix $A_j^{\pp}$ equal for all three spins in this calculation. Along any column or row, a hierarchy in the energies is evident. Also note the intrinsic symmetry of energies with respect to the checkerboard conjugate diagonal (see \zfr{fig4}). }
\zfl{fig12}
\end{figure}

 \vspace{-5mm}
\section{Hierarchy in energy gaps}
 \vspace{-1mm}
In this Appendix, we present details of the diagonalization of the Hamiltonian resulting in calculations of the energy gaps and their native hierarchy in the checkerboard $\mI_{k,\ell}$ (see \zsr{cascade}D). The procedure for the calculation of the $\vxe_{k,\ell}$ energy gaps is most simply illustrated considering first an $N{=}1$ system and evaluating the anti-crossing between states $\ket{0,\up}$ and $\ket{\R{+}1,\dw}$ corresponding to a ``joint" $e$-$n$ flip at checkerboard point $\mI_{1,1}$. There is no direct term in the Hamiltonian that can accomplish a dual spin-flip,  but as we described in \zsr{cascade}D, it can be constructed as a combined action of the $\{\xO_e,A_{1}^{\pp}\}$ terms. Indeed, the Hamiltonian $\mH$ in \zr{eq4} in the basis indicated (blue terms) has the form,

\beq
\mH - \mI_{1,1}(1)\T{1} = \begin{bNiceMatrix}[first-row,,first-col,last-col]
    & \ket{0,\up} & \ket{+1,\dw} & \ket{0,\dw} & \ket{+1,\up}       \\
 & 0   &   &    & \xO_e   & \ket{0,\up} \\
&    & 0   & \xO_e   & A_1^{\pp}  & \ket{+1,\dw} \\
&    & \xO_e   & -\xo_1^{(0)}   &    & \ket{0,\dw} \\
& \xO_e   &A_1^{\pp}   &    &  \xo_1^{(1)}   &\ket{+1,\up}  \\
\end{bNiceMatrix}
\zl{Hdiagonalize1}
\eeq
where we employ a $\xo_{\R{MW}}$ value in \zr{eq4} to be resonant with the checkerboard point $\mI_{1,1}(1)$. Clearly the states $\ket{0,{\up}}$ and $\ket{\R{+}1,{\dw}}$ are then degenerate, and while no direct coupling (off-diagonal term) exists between them, it can be constructed by first diagonalizing with respect to the $A_1^{\pp}$ term.  The new eigenstates then can be defined as,
\bea
\ket{\R{+}1,{\dw}}_A&=&\ket{\R{+}1,{\dw}}\cos\xt_A^{(1)} +\ket{\R{+}1,{\up}}\sin\xt_A^{(1)} \non\\
\ket{\R{+}1,{\up}}_A&=&-\ket{\R{+}1,{\dw}}\sin\xt_A^{(1)} +\ket{\R{+}1,{\up}}\cos\xt_A^{(1)}\non\\
\ket{0,{\dw}}_A&=&\ket{0,{\dw}}\: ;\:\ket{0,{\up}}_A=\ket{0,{\up}}
\zl{Abasis}
\eea
where $\tan\xt_A^{(1)} {=} A_1^{\pp}/\xo_1^{(1)}$, and we have used the fact that the hyperfine coupling is predominantly operational only in the $m_s{=}$+1 manifold. Restricting our attention to the degenerate two-level subspace in \zr{Hdiagonalize1} gives,
\beq
\mH - \mI_{1,1}(1)\T{1} \app \begin{bNiceMatrix}[first-row,,first-col,last-col]
   & \ket{0,\up}_A & \ket{1,\dw}_A   &   \\
 &0    &  \xO_e\sin\xt_A^{(1)}& \ket{0,\up}_A \\
&   \xO_e\sin\xt_A^{(1)}   &  \fr{1}{2}(A_1^{\pp})^2/\xo_{1}^{(1)}& \ket{1,\dw}_A \\
\end{bNiceMatrix}
\eeq

allowing us to read off the energy gap,
\beq
\vxe_{1,1}=  \lsb \lb\fr{1}{2}(A_1^{\pp})^2/\xo_{1}^{(1)}\rb^2 + (2\xO_e\sin\xt_A^{(1)})^2\rsb^{1/2}
\eeq
In the simple case that $ A_1^{\pp}{\ll}\xo_1^{(1)}$, this gives,
$
\vxe_{1,1}\app \fr{2\xO_e A_1^{\pp}}{\xo_L+A_1^{\pll}}$.  Indeed, this result could also have been predicted out of simple first-order perturbation theory arguments. A similar calculation for the energy gap $\vxe_{2,1}$ shows that a hierarchy in the $N{=}1$ checkerboard is operational, $\vxe_{1,1} {<} \vxe_{2,1} {=} \xO_e$. This is illustrated for instance in \zfr{fig8}.

Similarly, now consider the diagonalization of the Hamiltonian for the $N{=}2$ case. We begin the discussion by noting that in the basis indicated, the Hamiltonian has the form,
\begin{widetext}
\beq
\mH  = \begin{bNiceMatrix}[first-row,,first-col,last-col]
    & \ket{0,\up\up}  & \ket{\R{+}1,\dw\dw} & \ket{\R{+}1,\up\dw} & \ket{\R{+}1,\dw\up}&\ket{\R{+}1,\up\up}         \\
& 0 & & & & \xO_e &  \ket{0,\up\up}\\ 
& &0 &A_1^{\pp} & A_2^{\pp} & & \ket{\R{+}1,\dw\dw}\\ 
& &A_1^{\pp} &\xo_{1}^{(1)} & & & \ket{\R{+}1,\up\dw}\\ 
& &A_2^{\pp} & & \xo_{1}^{(2)}& & \ket{\R{+}1,\dw\up}\\ 
&\xO_e & & & & -(\xo_{1}^{(1)} + \xo_{1}^{(2)}) & \ket{\R{+}1,\up\up}  \\ 
 \end{bNiceMatrix}
\eeq
\end{widetext}
where there is an energy degeneracy between states $ \ket{0,\up\up}$ and  $\ket{\R{+}1,\dw\dw}$, but no direct coupling between them.  Once again, however, we can carry out a diagonalization with respect to the $\{ A_1^{\pp}, A_2^{\pp} \} $ terms, identifying a rotated basis similar to \zr{Abasis},
\bea
\ket{\R{+}1,{\dw\dw}}_A&=&\ket{\R{+}1,{\dw\dw}}\cos\xt_A^{(1)} +\ket{\R{+}1,{\up\dw}}\sin\xt_A^{(1)} \non\\
\ket{\R{+}1,{\up\up}}_A&=&-\ket{\R{+}1,{\up\dw}}\sin\xt_A^{(2)} +\ket{\R{+}1,{\up\up}}\cos\xt_A^{(2)}
\eea
allowing one to write the Hamiltonian matrix in the degenerate subspace,
\beq
\mH \app \begin{bNiceMatrix}[first-row,,first-col,last-col]
   & \ket{0,\up\up}_A & \ket{1,\dw\dw}_A   &   \\
 &0    &  2\xO_e\sin\xt_A^{(1)}\sin\xt_A^{(2)}& \ket{0,\up\up}_A \\
&  2 \xO_e\sin\xt_A^{(1)}\sin\xt_A^{(2)}   &  \fr{1}{2}(A_1^{\pp})^2/\xo_{1}^{(1)} + \fr{1}{2}(A_2^{\pp})^2/\xo_{2}^{(1)}& \ket{1,\dw\dw}_A \\
\end{bNiceMatrix}
\eeq
ultimately resulting in an energy gap that can approximately be written as,
\beq
\vxe_{1,1}\app 4\xO_e\sin\xt_A^{(1)}\sin\xt_A^{(2)}=  4\xO_e \lb\fr{A_1^{\pp}}{\xo_L+A_1^{\pll}}\rb \lb\fr{A_2^{\pp}}{\xo_L+A_2^{\pll}}\rb\:,\non
\eeq
Clearly a hierarchy in the energy gaps is then evident, since $\vxe_{2,1}{\app} 2\xO_e\sin\xt_A^{(1)}$ and $\vxe_{3,1}{\app} \xO_e$, and if the transverse hyperfine terms are approximately the same order of magnitude,
$\vxe_{1,1}{<} \vxe_{2,1}{<}\vxe_{3,1}$, 
generally decreasing with the number of successive nuclear flips required between the two states at each anti-crossing. 
\zfr{fig12} displays the energy gaps $\vxe_{k,\ell}$ for an $N{=}3$ system — the heights of the bars here are the size of the energy gaps, while their positions correspond to those on the checkerboard. We have chosen $A_j^{(1)}$ equal here to highlight the hierarchy just coming from the position of the anti-crossing on the checkerboard. 

\bibliography{Galton_model_FINAL2.bbl}
\end{document}

%% file: Commands3.tex
\newcommand{\xb}{\beta}

\newcommand{\vxe}{\varepsilon}

\newcommand{\xg}{\gamma}
\newcommand{\xt}{\theta}

\newcommand{\xo}{\omega}
\newcommand{\xph}{\phi}

\newcommand{\pp}{\perp}
\newcommand{\app}{\approx}
\newcommand{\dw}{\downarrow}
\newcommand{\up}{\uparrow}

\newcommand{\Cs}{{}^{13}\R{C}}

\newcommand{\mC}[0]{\mathcal{C}}
\newcommand{\mI}[0]{\mathcal{I}}

\newcommand{\mB}[0]{\mathcal{B}}

\newcommand{\dxo}[0]{\dot\omega}

\newcommand{\ltrt}{\leftrightarrow}

\newcommand{\xD}{\Delta}

\newcommand{\xO}{\Omega}

\newcommand{\mP}[0]{\mathcal{P}}
\newcommand{\mO}[0]{\mathcal{O}}

\newcommand{\fr}[2]{\frac{#1}{#2}}

\newcommand{\mH}[0]{\mathcal{H}}

\newcommand{\mS}[0]{\mathcal{S}}

\newcommand{\rt}{\rightarrow}

\newcommand{\beq}{\begin{equation}}
\newcommand{\eeq}{\end{equation}}
                  
\newcommand{\benum}{\begin{enumerate}}
\newcommand{\eenum}{\end{enumerate}}
                    
\newcommand{\bit}{\begin{itemize}}
\newcommand{\eit}{\end{itemize}}

\newcommand{\bea}{\begin{eqnarray}}
\newcommand{\eea}{\end{eqnarray}}

\newcommand{\bev}{\begin{verbatim}}
\newcommand{\eev}{\end{verbatim}}

\newcommand{\non}{\nonumber}

\newcommand{\qt}{\tau}

\newcommand{\lb}{\left(}
\newcommand{\rb}{\right)}
\newcommand{\lsb}{\left[}
\newcommand{\rsb}{\right]}

\newcommand{\pll}{\parallel}

\newcommand{\T}[1]{\textbf{#1}}
\newcommand{\I}[1]{\textit{#1}}
\newcommand{\R}[1]{\textrm{#1}}

\newcommand{\zl}[1]{\label{eqn:#1}}
\newcommand{\zr}[1]{Eq.\,(\ref{eqn:#1})}
\newcommand{\zfl}[1]{\protect\label{fig:#1}}
\newcommand{\zfr}[1]{\figurename\,\ref{fig:#1}}

\newcommand{\zsl}[1]{\label{sec:#1}}
\newcommand{\zsr}[1]{Sec. \ref{sec:#1}}

\newcommand{\ket}[1]{\left\vert{#1}\right\rangle}

\newcommand{\expec}[1]{\left\langle #1\right\rangle}

\newcommand{\mat}[4]{\left(\begin{array}{cc}{#1}&{#2}\\
{#3}&{#4}\end{array}\right)}

\newcommand{\ba}{\left\{ \begin{array}{lr}}
\newcommand{\ea}{\end{array}\right.}

\newcommand{\blist}[1]{
 \begin{list}{#1}
 \begin{align}
	 arrow
 \end{align}
 $\checkmark\star
  { \setlength{\itemsep}{3pt}
     \setlength{\parsep}{2pt}
     \setlength{\topsep}{3pt}
     \setlength{\partopsep}{0pt}
     \setlength{\leftmargin}{1em}
     \setlength{\labelwidth}{1em}
     \setlength{\labelsep}{0.5em} } }
\newcommand{\elist}{
  \end{list}  }

\DeclareMathSymbol{\vartheta}{\mathalpha}{letters}{"12}
\DeclareMathSymbol{\theta}{\mathalpha}{letters}{"23}
\DeclareMathSymbol{\phi}{\mathalpha}{letters}{"27}
\DeclareMathSymbol{\varphi}{\mathalpha}{letters}{"1E}

\newcommand{\bef}
{
\begin{figure}[htbp]
\centering
}

\newcommand{\eef}{\end{figure}}
